\makeatletter
\declare@file@substitution{revtex4-1.cls}{revtex4-2.cls}
\makeatother

\documentclass{aastex631}

\newcommand{\kms}{~km~s$^{-1}$}
\newcommand{\angs}{~\AA~}
\newcommand{\mangs}{~m\AA~}

\begin{document}

\title{On the Response of the Transition Region and the Corona to Rapid Excursions in the Chromosphere}

\author[0009-0002-0549-1446]{Ravi Chaurasiya}
\affiliation{Udaipur Solar Observatory, Physical Research Laboratory, Udaipur-313001, India} \affiliation{Indian Institute of Technology, Gandhinagar, Gujarat-382355, India} 
\author[0000-0001-5802-7677]{A. Raja Bayanna}
\affiliation{Udaipur Solar Observatory, Physical Research Laboratory, Udaipur-313001, India}
\author[0000-0001-5963-8293]{R. E. Louis}
\affiliation{Udaipur Solar Observatory, Physical Research Laboratory, Udaipur-313001, India}
\author[0000-0003-4747-4329]{T.M.D. Pereira}
\affiliation{Rosseland Centre for Solar Physics, University of Oslo, PO Box 1029, Blindern 0315, Oslo, Norway}
\affiliation{Institute of Theoretical Astrophysics, University of Oslo, PO Box 1029, Blindern 0315, Oslo, Norway}

\author[0000-0002-9370-2591]{S. K. Mathew}
\affiliation{Udaipur Solar Observatory, Physical Research Laboratory, Udaipur-313001, India}

\begin{abstract}

Spicules are the thin hair/grass-like structures that are prominently observed at the chromospheric solar limb. It is believed that fibrils and rapid blueshifted and redshifted  excursions (RBEs and RREs; collectively referred to as REs) correspond to on-disk counterparts of type I spicules and type II spicules, respectively. Our investigation focuses on observing the response of these REs alongside similar spectral features in the chromosphere, transition Region (TR), and corona, utilizing space-time plots derived from coordinated observations from Swedish 1 m Solar Telescope/H$\alpha$, Interface Region Imaging Spectrograph (IRIS), and Solar Dynamics Observatory. Our analysis reveals upflowing REs, promptly reaching temperatures characteristic of the TR and corona, indicating a multi-thermal nature. Similarly, downflowing features exhibiting similar spectral signatures over the disk display plasma motion from the corona to chromospheric temperatures, demonstrating a multithermal nature. In addition to distinct upflows and downflows, we observe sequential upflow and downflow along the same path, depicting a distinctive parabolic trajectory in space-time plots of observations sampling TR and various coronal passbands. Similar to isolated upflows and downflows, these REs also exhibit a multi-thermal nature throughout their trajectory. Furthermore, our results reveal a more intricate motion of the REs in which both upflow and downflow coexist at the same spatial location. On a different note, our analysis, utilizing coordinated IRIS spectral observations, shows spatio-temporal redshifts/downflows in both the TR and chromosphere, suggesting that at least subsets of the strong redshifts/downflows observed in TR temperature spectra result from the return from the upper atmosphere flow of plasma in the form of bundles of spicules or features exhibiting similar spectra.

\end{abstract}


\section{Introduction}

The chromosphere is one of the layers of the solar atmosphere, sandwiched between the photosphere and the corona, which have drastically different physical properties and where magnetic pressure starts to dominate over gas pressure due to decreasing density and increase in temperature. It is dominated by many features such as filaments, umbral flashes, dynamic fibrils, mottles, spicules, and etc. These features contribute to the chromosphere's rapid dynamical changes on timescales spanning from seconds to minutes, requiring high spatio-temporal observations. This layer plays an important role in mass loading and heating the upper atmosphere, as all of the nonthermal energy responsible for mechanisms that contribute to coronal heating and driving solar wind propagates via the chromosphere \citep{2019ARA&A..57..189C} [and references therein].

It is believed that spicules, which are one of the most abundant features in the chromosphere, contribute to coronal heating and supply of mass to the solar wind \citep{1968SoPh....3..367B, 1978SoPh...57...49P, 1982ApJ...255..743A, 2019ARA&A..57..189C}. Spicules are ubiquitously seen as bright elongated grass-like structures over the solar limb in the prominent chromospheric spectral line, such as $H\alpha$ and Ca II 8542 \angs \citep{secchi1877soleil, 1945ApJ...101..136R, 1968SoPh....3..367B, 1972ARA&A..10...73B, 2000SoPh..196...79S, 2012SSRv..169..181T}. In contrast, the same structures appear as absorption features and are seen as thin, elongated, thread-like, and highly dynamic structures over the disk. These spicules may be seen all across the solar surface, and it is estimated that more than a million can always be seen on the solar disk, with small variations in size \citep{2010ApJ...719..469J}. Only one kind of spicules existed until the discovery of a more elusive, but energetic category of spicules by \cite{de2007tale}  using observations from the Solar Optical Telescope (SOT; \citep{HinodeSOT}) on board Hinode \citep{2007SoPh..243....3K}. This led to the classification of spicules as type I and type IIs.

Type-I spicules are dynamic fibrils observed in active regions (ARs) or quiet Sun mottles (QS), and they typically manifest themselves close to strong magnetic fields \citep{2004Natur.430..536D}. These spicules follow parabolic trajectories in  H$\alpha$ line core distance-time diagrams with typical upward and downward velocities of the order 10--40\kms having lifetimes between 3--5 min and showing quasi-periodicities of almost the same time periods \citep{2007ApJ...660L.169R}. \cite{2004Natur.430..536D} showed that type I spicules are generated due to the leakage of photospheric magnetoacoustic oscillations into the solar chromosphere, which was later firmly established by \cite{2006ApJ...647L..73H} and \cite{2007ApJ...660L.169R} who found remarkable similarities between the observations and their advanced numerical simulations. On the other hand, type II spicules are much more dynamic with rapid sideways motions, high apparent speeds and lifetimes of the range 80--300\kms and ~ 1-3 minutes, respectively \citep{de2007tale, 2012ApJ...759...18P, 2016ApJ...824...65P}. While the origin of type I spicules is well established, the origin of type II spicule is still unclear \citep{2019ARA&A..57..189C}. 

It was difficult to trace and deduce the properties of type II spicules from limb observations due to the superposition of a number of features along the line of sight. Thus, finding the on-disk counterpart of the type II spicules was important. The first spectral signature of the on-disk counterparts of type-II spicules was observed by \cite{2008ApJ...679L.167L} who found short-lived ($\sim$ 40s), high-speed events with speeds of 15-30\kms that appeared as sudden rapid excursions in the blue wings of Ca II 8542\angs without any subsequent redshifts and hence named them rapid blue shifted excursions (RBEs). Later, using high resolution on disk images and spectra in spectral lines  H$\alpha$, Ca II 8542\angs and Ca II K, \cite{2009ApJ...705..272R, 2012ApJ...752..108S, 2016ApJ...824...65P} firmly established that RBEs were indeed the on-disk counterparts of the type-II spicules. 

According to \cite{2012ApJ...752L..12D}, type-II spicules exhibit torsional motions, which account for the spicular bushes seen in red-wing images of chromospheric spectral lines that morphologically resemble their blue wings counterparts. Following this, \cite{2013ApJ...769...44S} also found the red wing counterparts of RBEs in H$\alpha$ and Ca II 8542\angs lines with comparable morphologies to the former which were termed rapid red-shifted excursions (RREs). \cite{2015ApJ...802...26K, 2015ApJ...799L...3R} verified their presence with high-resolution on-disk observations, both in the chromosphere and the TR and concluded that RREs, although less common than RBEs, were manifestations of the same phenomenon with identical statistical properties. These RREs were detected due to the complicated simultaneous action of swaying, tortional and field aligned motion, which caused a net red shift when viewed on the disk. Thus, RBEs can switch to RREs and vice-versa depending upon the angle between the LOS and transverse motion of the structures. Time series images of  RBEs and RREs suggests that both seem to be generated near strong magnetic regions and move upward as they evolve \citep{2009ApJ...705..272R, 2009ApJ...707..524M, 2013ApJ...769...44S, 2014Sci...346D.315D, 2015ApJ...802...26K}. The complex twisting and swaying found in type-II spicules as they propagate along the chromospheric magnetic field lines are the representative of outward propagating Alfvénic waves, which can be of the order of several hundred \kms and can cause heating in the hotter TR lines \citep{2014Sci...346D.315D}. 
However, a closer look at the time series images at the red wings of chromospheric lines, such as H$\alpha$ and Ca II 8542 \AA \space lines, shows another class of absorption features whose spectral signature and morphology are identical to traditional RBEs and RREs; however, their direction of motion is opposite to traditional RBEs and RREs. Recently, \cite{2021A&A...647A.147B} have referred to them as Downflowing RREs as they are rapid downflows in the solar chromosphere in the form of spicules and speculated that these downflowing RREs are the chromospheric counterparts of observed redshift in the transition region. 

\subsection{Upflows in the Solar atmosphere}

Spicules were considered as a potential candidate to contribute to the mass supply to the corona/solar wind and coronal heating \citep{1968SoPh....3..367B, 1978SoPh...57...49P, 1982ApJ...255..743A}. However, earlier studies \citep{1983ApJ...267..825W, 1992str..book.....M} ruled out the role of spicules in balancing the mass and energy in the corona as they could not find signatures of spicules at coronal temperatures. The discovery of type~II spicules (\cite{de2007tale}) led to interesting results in understanding the role of spicules in mass supply to corona. \cite{2011Sci...331...55D} used coordinated observations from the Hinode and SDO missions to provide one of the first conclusive findings of heated spicules responsible for feeding hot plasma to the solar corona. Later, \cite{2014ApJ...792L..15P} using observation from Hinode and IRIS, clearly established that after fading from the Ca II H filter, spicules continue to evolve in passbands that are sensitive to higher temperatures, such as Mg II and Si IV, sampling the upper chromosphere and TR. In addition, many studies \citep{2011Sci...331...55D, 2014ApJ...792L..15P, 2015ApJ...799L...3R, 2016ApJ...820..124H, 2016ApJ...830..133K, 2019Sci...366..890S} have shown that type II spicules are heated to TR and corona temperatures as they propagate beyond the chromosphere both in quite sun and active regions.  Although the role of spicules is well-established in the dynamics of the transition region, their impact on the corona and solar wind remains a subject of debate \citep{2019ARA&A..57..189C}. 

\subsection{Downflows in the Solar atmosphere}

Downflows are considered to play an important role in understanding the mass and energy balance of the solar atmosphere. Observations over the decades have shown the predominance of the downflows in the TR spectral lines \citep{1976ApJ...205L.177D, 1981ApJ...249..720S, 1982SoPh...77...77D, 2011A&A...534A..90D}. High-speed downflows observed in TRs can sometimes last several hours to several days \citep{1981ApJ...249..720S}. According to these studies, the average line profiles formed in the transition region are up to 10-15 \kms redshifted, indicating the presence of plasma flows or wave motion with amplitudes that are significant fractions of the speed of sound.

However, on-disk observations from the Solar and Heliospheric Observatory (SOHO) and the Hinode Extreme Ultraviolet Imaging Spectrograph (EIS, \cite{2007SoPh..243...19C}) have shown that the spectral lines which form beyond 1 MK (coronal lines) show a net upflow or blueshift on the order of a few \kms \citep{1999ApJ...516..490P, 1999ApJ...522.1148P, 2009ApJ...701L...1D, 2012ApJ...749...60M}. This finding suggests that the net Doppler shift of these spectral lines may be temperature-dependent, changing from redshifts to blueshifts. According to \cite{2012ApJ...749...60M}, this conversion happens around 1 MK. Thus, the above two findings led to several hypotheses to explain the observed shifts in TR and coronal line from both perspectives (observations and numerical simulations). These theories include the return of the previously heated spicular material \citep{1977A&A....55..305P, 1982ApJ...255..743A}, fast episodic heating at the base of the corona \citep{hansteen2010redshifts}, downward propagating pressure disturbances \citep{zacharias2018disentangling}, and downward propagating acoustic waves produced by nanoflares in the corona \citep{1993ApJ...402..741H}. However, no conclusive agreement has yet been reached.

Due to their ubiquitous presence, spicules may be a key factor in explaining these Doppler shifts in the higher atmosphere. \cite{1977A&A....55..305P} were the first persons to propose that the downflowing spicular materials could explain the observed redshifts in the transition region lines because the average upward mass flux carried by spicules was about 100 times more than what is required to contribute to the solar wind. Thus, after cooling down, the remaining heated spicular material falls back to the chromosphere via the transition region. Later, this scenario was validated by  \cite{1982ApJ...255..743A} and \cite{1984ApJ...287..412A} based on their thorough modelling work. These studies lacked observational support, and more sophisticated theoretical models have failed to replicate these first results \citep{1987ApJ...319..465M}. However, recent advancements in observations from high-resolution space-based telescopes in the Extreme Ultraviolet (EUV) and Ultraviolent (UV) passbands have sparked intense curiosity about the role of spicules in higher solar atmospheres. Strong observational evidence of lower coronal and transition region flows linked to spicules was found by \cite{2012ApJ...749...60M}. They found evidence of downflowing patterns in the comparatively cold emission channels in the lower solar corona, which were thought to be the return of heated spicular plasma.

\begin{figure}
\centering
\includegraphics[width=0.475\textwidth]{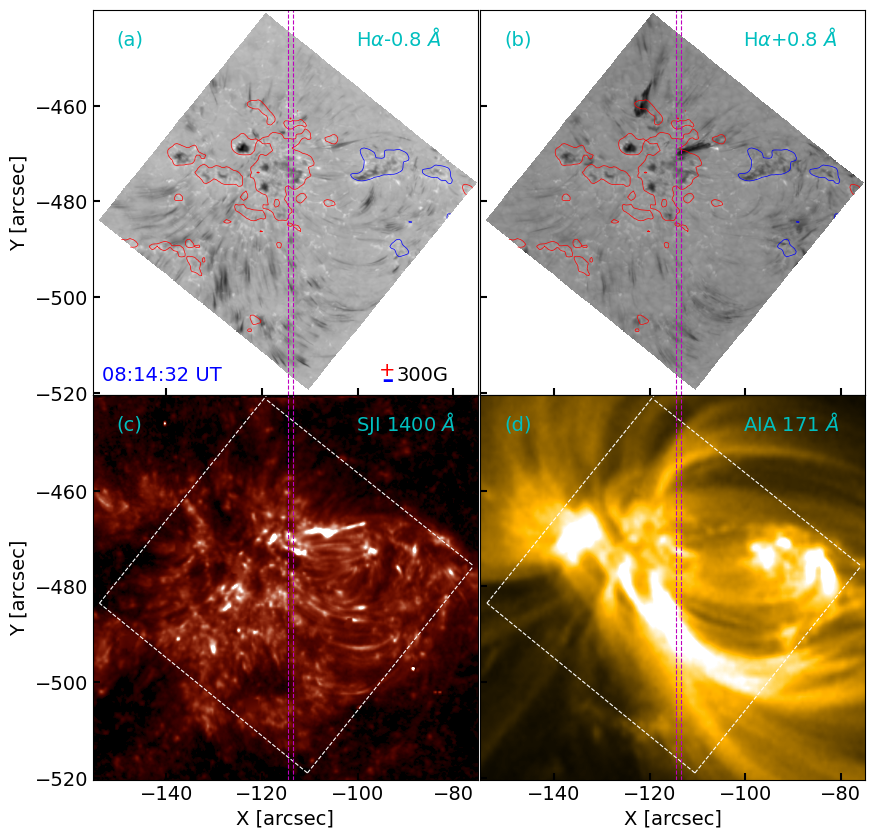}
\caption{Observations and spatial extent of the data used in this work. Panel (a) and (b) are the line scan image at blue wing (H$\alpha$-0.8\angs) and red wing (H$\alpha$+0.8\angs) of H$\alpha$ spectral line, respectively. Contours of the line-of-sight magnetic field of +300~G and -300~G are overlaid on the images. The magnetogram data was obtained from SDO/HMI. Panel (c) and (d) are the coaligned image of IRIS SJI 1400\angs and SDO/AIA 171\angs. The spatial extent of the IRIS slit is represented by dashed vertical lines in purple.}
\label{Fig1}
\end{figure}

Despite the predominance of downflows observed in the TR, \cite{1997SoPh..175..349B} and \cite{1999ApJ...522.1148P} showed that most high-speed quiet sun TR downflows usually vanish at chromospheric temperatures. \cite{2018ApJ...859..158S} did a statistical analysis of TR downflows in 48 sunspots and found that a maximum of 24 of them show signatures in the upper chromospheric spectral lines such as Mg II k, whereas the rest are limited to TR. Since sunspots and ARs occupy limited regions on the solar surface it is still unclear what is happening to the other downflows seen predominantly in the TR spectral line. It has been speculated that the absence of a signature of downflows in the chromosphere could be due to the breakdown of these downflows as soon as they reach the chromospheric temperature, as the plasma density is several orders of magnitude higher in the chromosphere than the TR and the corona. 

In this work, using coordinated observation from ground based and space based observatories, we study the response of these RBE/RREs alongside those chromospheric on-disk absorption features which show similar spectra as RBE/RREs, as categorised by $k-$means clustering (as discussed in section \ref{sec3}), in TR and coronal passbands. 

The rest of paper is structured as follows. Section \ref{sec2} outlines the observations and data alignment process. Section \ref{sec3} elaborates on the methodology employed to detect RBE/RREs from the SST observations. Results are presented in Section \ref{sec4}, followed by a discussion in Section \ref{sec5}. Finally, we conclude the paper with a summary in Section \ref{sec6}.

\section{Observations} \label{sec2}

We analyzed a data set consisting of NOAA AR 12775 observed on 11 October 2020 from 08:04 UT to 09:15 UT by the Swedish 1-m Solar Telescope (SST; \cite{2003SPIE.4853..341S}) along with co-spatial and co-temporal data sets from Interface Region Imaging Spectrograph (IRIS; \cite{2014SoPh..289.2733D}) Solar Dynamic Observatory (SDO, \cite{2012SoPh..275....3P})/Atmospheric Imaging Assembly (AIA, \cite{2012SoPh..275...17L}) and SDO/Helioseismic Magnetic Imager (HMI,\space \cite{2012SoPh..275..229S}). The heliocentric coordinates of the target AR were  ($-$114\arcsec, $-$479\arcsec)
, with the corresponding observing angle µ = cos $\theta$ = 0.86 (where $\theta$ represents the heliocentric angle). The specified AR represented a network region with moderate magnetic field and contained multiple pores, as illustrated in Figure \ref{Fig1}. Further information regarding the data from various instrumentation facilities is provided below.

\subsection{Swedish Solar Telescope (SST)}
 The SST data set consists of imaging spectroscopic observations of H$\alpha$, H$\beta$, and Ca II 8542\angs. H$\alpha$, Ca II 8542\angs observations were obtained using the CRisp Imaging SpectroPolarimeter (CRISP, \cite{scharmer2008crisp}) while the H$_\beta$ observations were obtained from the CHROmospheric Imaging Spectrometer (CHROMIS, \cite{scharmer2017solarnet}). In this work, we have analyzed only the H$\alpha$ observations. The temporal duration of the H$\alpha$ data was nearly 66 minutes consisting of two data cubes, the first starting from 08:04 UT to  08:35 UT  and the second starting from 08:39 to 09:15 UT with temporal cadence of 34.252 s. The field-of-view (FoV) of the observation was approximately 60\arcsec$\times$60\arcsec with a plate scale of 0\arcsec.0591 pixel$^{-1}$. The H$\alpha$ line was scanned at 17 wavelength positions with a spectral coverage of $\pm$2\angs. The atmospheric seeing effects in the data are corrected by SST adaptive optics system \citep{2019A&A...626A..55S} and further processed using the Multi-Object Multi-Frame Blind Deconvolution (MOMFBD, \cite{van2005solar}) image restoration technique. This data-set data SST were made available by the SST team publicly{\footnote{https://dubshen.astro.su.se/sst$\_$archive/search}}.

\begin{figure}
\centering
\includegraphics[width=0.45\textwidth]{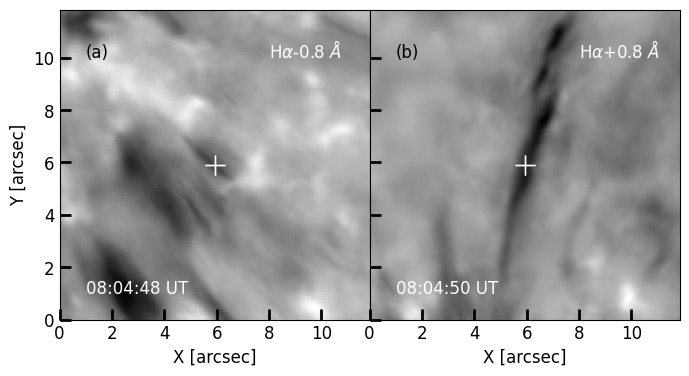}\\
\includegraphics[width=0.485\textwidth]{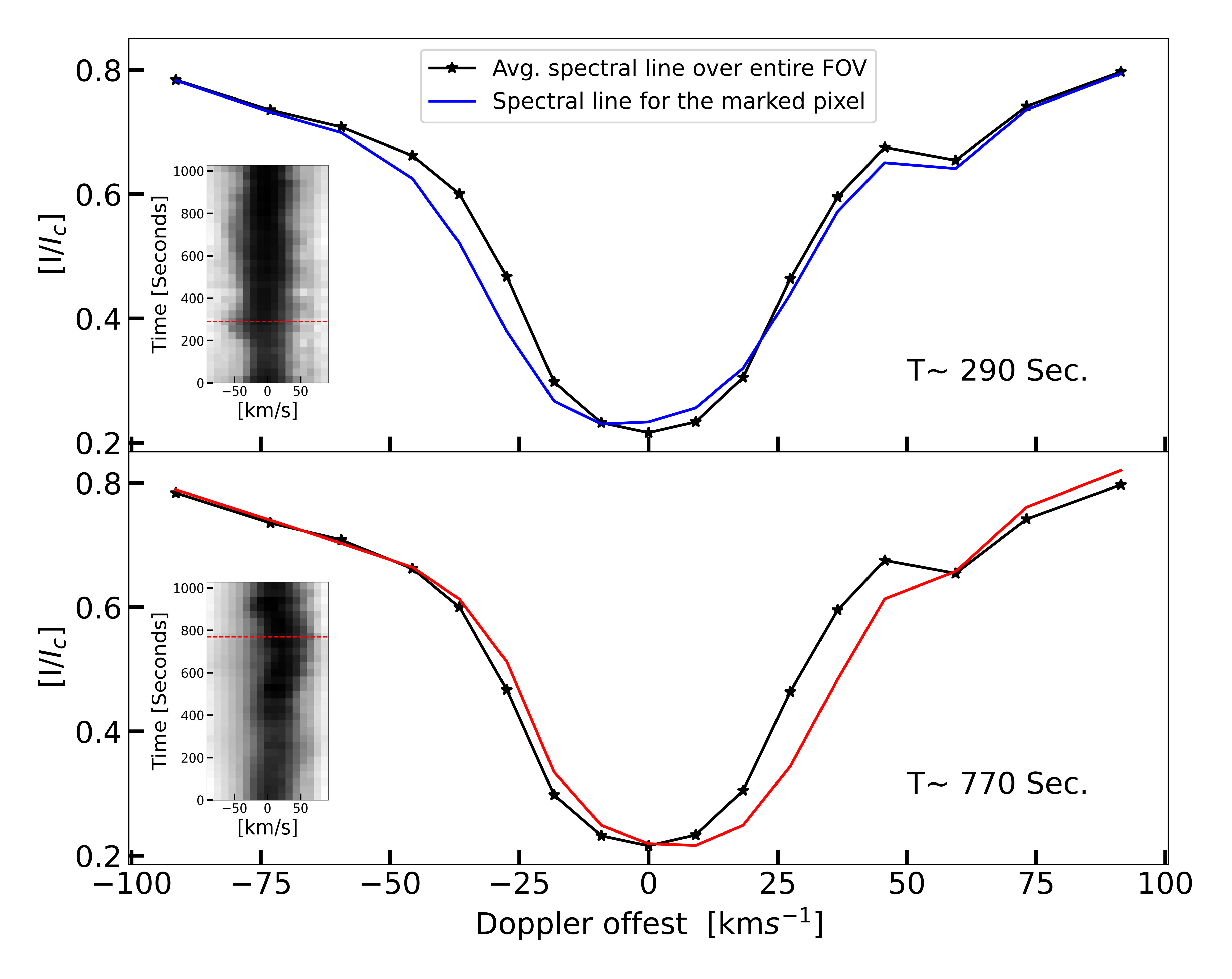}
\caption{Panel (a) and (b) are part of the whole FoV of the SST/H$\alpha$ observations at blue wing (H$\alpha$-0.8\angs) and red wing (H$\alpha$+0.8\angs), respectively. The corresponding middle and bottom panels show the H$\alpha$ spectral line profiles for the marked pixel with \textbf{+} sign at a particular instant. Inset images show the $\lambda$-t maps of marked pixels indicating RBE and RRE.}  
\label{Fig2}
\end{figure}

\subsection{Interface Region Imaging Spectrograph (IRIS)}

The corresponding coordinated IRIS\footnote{IRIS OBSID: 3630108417} data sets, which include spectra and Slit-Jaw Images (SJIs), were obtained for the given coordinated SST and IRIS campaign. The SJIs were taken in three wavelength windows, which include Si IV 1400\angs, Mg II k 2796\angs and Mg II h wing 2832\angs with an identical plate scale of 0.3327$"$ and time cadence of 21.88 s, 18.22 s and 109.35 s respectively. Here, we have used Si~IV and Mg~II~k observations only. Apart from SJIs, IRIS ran in a large, dense 4-step raster (step size of 0.35$"$). The step cadence was 9.2 s, giving the raster cadence of about 32s. It took spectral observations in C II 1336\angs ($\sim$ 2 $\times$ 10$^{4}$ K ), Si IV 1394\angs ($\sim$ 10$^{4.8}$ K), Si IV 1403\angs ($\sim$ 10$^{4.8}$ K), and Mg II k 2796\angs ($\sim$ 10$^{4.2}$ K). The spatial sampling along the slit was 0.3327$"$, and the spectral sampling was 25.3~\mangs and 50.7\mangs for near UV (Mg II k) and far UV (Si IV and C II) windows, respectively. In this work, we used only Si IV 1394\angs spectral observation. The spatially averaged spectra over the large region of neutral and singly ionised species formed in the lower chromosphere is generally considered to be at rest \citep{1991ApJ...372..710H}. Here, we have used O I 1356\angs spectra for calibrating Si IV 1394\angs spectra.

To coalign the images from the two instruments (SST and IRIS-SJI), the IRIS SJIs were expanded by bilinear interpolation to SST/H$\alpha$ plate scale, followed by the cross-correlation between the almost simultaneous IRIS 2832\angs SJI image and the photospheric wing image from SST/H$\alpha$. Finally, the same range of pixels was extracted from all the wavelength channels of SJIs to get 80" $\times$ 80" FoV. To ensure the proper coalignment, different contrasting features were taken and cross-correlated again, making the offset as minimal as possible. 

\subsection{Solar Dynamics Observatory (SDO)}
The corresponding registered coordinated cutouts were downloaded using SDO cutout sequence, which includes EUV data from SDO/AIA and continuum image and magnetograms from SDO/HMI. The AIA--UV (1600\angs, 1700\angs) and AIA--EUV ( 171 \angs, 193\angs, 211\angs and 304 \angs) were having the approximately identical plate scale of 0$\arcsec$.6 pixel$^{-1}$  with different time cadence of 12s and 24s respectively. The HMI images (LOS magnetogram and Continnum) were having a approximately identical cadence  and plate scale of 45s and 0$\arcsec$.5 pixel$^{-1}$ respectively. To coalign the images from the two instruments and extract the SST FoV from the SDO cutouts, the SDO cutouts were expanded by bilinear interpolation to SST/H$\alpha$ plate scale in the first place. Coaligning the AIA images directly with the SST observations was difficult since the features are completely different at these heights. Hence we used AIA 1700\angs co-temporal images to co-align with the SST/H$\alpha$ continumm images and obtained AIA cutouts with FoV of~80" $\times$ 80" as SST observations. Here also, to ensure the proper coalignment, different contrasting features were taken and cross-correlated again. The same procedure is adopted to align the HMI continuum images. The procedure used while dealing with the SJIs is not used with SDO cutouts, as there is some offset between the plate scale of AIA EUV and UV channels. 

Figure \ref{Fig1} shows the line-scan images of SST/H$\alpha$ at  H$\alpha$-0.8\angs and at H$\alpha$+0.8\angs  along with coaligned IRIS SJI 1400\angs and AIA 171\angs images. The dashed vertical lines in purple represents the IRIS slit coverage over the FoV.

\section{Categorising RBEs and RREs} \label{sec3}

The SST/H$\alpha$ data consists of RBEs and RREs seen predominantly as the absorption features in the wings of the H$\alpha$ spectral line. These REs move upward and downward, as evident in the animation of time series images. Since these REs possess velocity, they show up as broadening in the H$\alpha$ with respect to the average spectral line profile in that pixel. The H$\alpha$ spectral line of the marked pixel with respect to the average spectral line profile of the whole FoV is shown in Figure~\ref{Fig2}. The $\lambda$-time maps of the marked pixel revealing the rapid blue and rapid red excursions are shown in the inset.

\begin{figure*}
\centering
\includegraphics[width=0.9\textwidth]{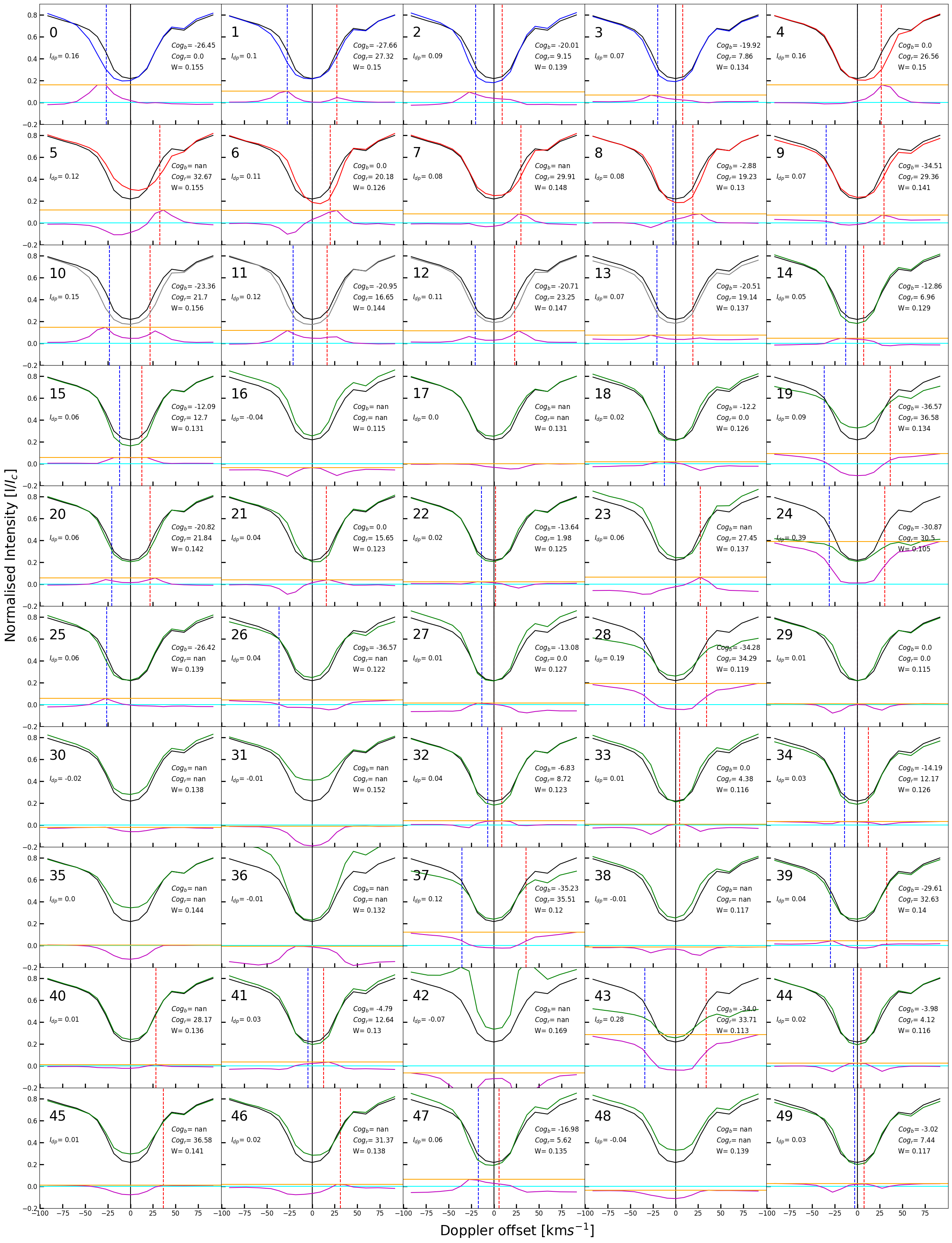}
\caption{The panel above displays 50 H$\alpha$ RPs (Representative Profiles) resulting from $k-$means clustering applied to the entire FoV). These RPs can be categorised into three groups: RPs(0--3) blue excursion, RPs (4--9) red excursion, and RPs (10--12) to excursion on either side. Additionally, other RPs correspond to various features on the solar disk, including areas with high magnetic fields and quiet sun regions. The black H$\alpha$ spectral line profile represents the average profile and serves as our reference profile. The magenta curve illustrates the variance between the RP and the reference profile. The horizontal orange and cyan lines indicate the points of maximum difference between the RP and the reference profile and the zero-normalized intensity value, respectively. The red and blue vertical lines and black signify the centre of gravity positions for the positive difference profile in the red, and blue wings of the H$\alpha$ spectral line and the line centre of the reference profile. The values of FWHM (W[nm]), the maximum difference between the RP and the reference line profile (I$_{dp}$), cog$_r$ [km s$^{-1}$] , and cog$_b$ [km s$^{-1}$], are provided in each RP plot.}
\label{Fig3}
\end{figure*}

In order to identify RBEs and RREs in the entire FoV, we have employed a $k-$mean clustering method for identifying RBEs and RREs, similar to these studies (\cite{2011A&A...530A..14V,2019ApJ...875L..18S,bose2019characterization,2024AdSpR..73.3256L}). The $k-$means clustering works by grouping or clustering observed line profiles into several groups or clusters on the basis of the spectral shape. Each line profile is indexed to the cluster center closest to it. One significant advantage of employing this technique is that it relies on the complete spectral signature of various features within the FoV, rather than solely considering their appearance at a specific wavelength. \cite{2016ApJ...824...65P} emphasized the significance of considering H$\alpha$ line widths when detecting and analysing spicules. Their findings revealed that relying solely on statistical properties measured from H$\alpha$ line wing images at specific wavelength positions could lead to an underestimation of the extracted physical quantities. This is because spicules exhibit a wide range of Doppler offsets during their evolution. In such cases, line width maps prove to be a more robust tool for spicule detection. Typically, both RBEs, RREs and downflowing RREs exhibit large Doppler offsets and enhanced line widths. The fastest spicules even display broader line profiles. This emphasises the importance of considering the Doppler offset and line width as crucial properties in spicule detection. The $k-$mean clustering approach facilitates efficient detection and characterization of these features. We determined the optimal number of cluster centres using the elbow method \citep{scikit-learn}, leading us to select 50 clusters. Every cluster possesses a representative profile (RP) that is the mean of all profiles within that cluster. All the 50 H$\alpha$ representative profiles (RPs) generated through the $k-$means clustering process are shown in Figure \ref{Fig3}.

\begin{figure}
\centering
\includegraphics[width=0.475\textwidth]{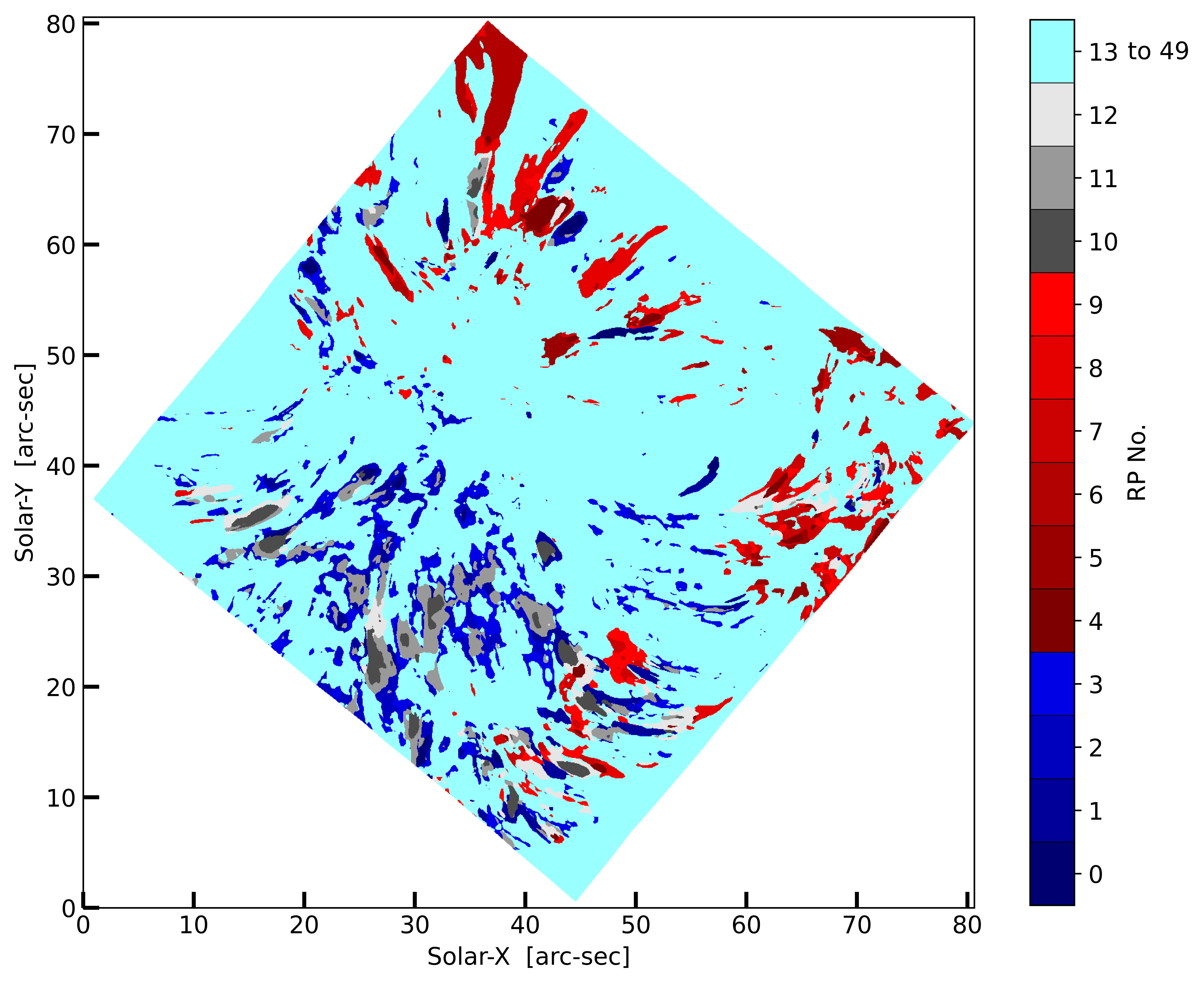}
\caption{Spatial distribution of RBEs (blue), RREs (red), dual flow (gray), and the remaining RPs (light blue) in the FOV. The color bar shows the RP number (refer Figure~\ref{Fig3}).}
\label{Fig4}
\end{figure}

Our approach to identifying spicules begins with the computation of differential profiles, by determining the difference between the average H$\alpha$ profile and the representative profile (RP). We computed the centre-of-gravity (COG) of the difference profiles (depicted by magenta in Figure \ref{Fig3}) separately for the red and blue parts from the line centre using only positive segments of the difference profiles depicted by the red and blue vertical lines. The combination of elevated values for the COG and the positive peak within the difference profile is an effective criterion for identifying enhanced absorption within the H$\alpha$ wings. We applied three specific criteria to identify RPs associated with RBEs and Rapid RREs. These RPs should: (1) exhibit a minimum COG Doppler offset (in either the red or blue wing) of 18 km s$^{-1}$, (2) have a maximum difference between the RP and the reference line profile (I$_{dp}$) $\ge$ 0.066 and (3) have line widths $\ge$ 1.25 \angs. Applying the aforementioned criteria we extracted 13 RPs in our data set in which clusters 0--3 belong to RBEs, 4--9 belongs to RREs (refer Figure \ref{Fig3}). It is evident that some of the RPs (10--12) show excursion on either wing. These RPs form the basis of the RBE/RREs with relatively low LOS velocity and are sometimes associated with dual flow on the disk and can be seen as absorption features on either wing of the H$\alpha$ spectral images. Although RP number 13 satisfies the aforementioned threshold criteria, it lacks excursions on either wing, unlike RP numbers 0--12. RP-- 13 exhibits a profile nearly parallel to the reference profile and is not associated with spicular dynamics when observed in the H$\alpha$ wings. It is to be noted that not all the RPs detected using the above threshold criteria are spicules; some regions have spectral signatures similar to those but do not show elongated structures. To visualise these RPs in our data set, we constructed a 2D map (Figure~\ref{Fig4}) of those pixels corresponding to these selected RPs. It is clear from Figure~\ref{Fig4} that most of the features in the FoV consist of RPs with different spectral characteristics (RBE, RRE and dual excursion). 

\section{Impact of RBEs and RREs in the Transition Region and coronal passbands} \label{sec4}

To understand the response of the upper solar atmosphere to the detected RBE/RREs or similar spectral features through $k-$means clustering in the chromosphere, we have constructed the space-time $(x-t)$ maps that show (a) only upflows, (b) only downflows (c) upflow and downflow along the same path one after the other and (d) dual flows along the same path simultaneously. The $x-t$ maps show the projected motion of the absorption/emission features in the sky plane that makes difficult to determine whether the feature is moving upward/downward in the solar atmosphere. Here, we classify these moving features as upflow (downflow) based on the wing images, which serve as a proxy for the doppler shift and motion of the features away from (towards) the magnetic flux concentration, as observed in the associated animations. These maps are constructed using the observations that sample the solar atmosphere at different heights corresponding to different temperature regimes. SJI 1400\angs images are used to construct $x-t$ map for the TR while AIA 171\angs, AIA 193\angs and AIA~211\angs images are used to construct similar $x-t$ maps for the corona. 

The virtual slit/box are drawn in such a way that it align well with the RBEs and/or RREs (collectively referred to as Rapid excursions (REs)). Instead of relying on a 1D virtual line, we employed a virtual 2D box/slit to account for potential alignment errors or any actual shifts in the observed features at multiple heights within the solar atmosphere. The starting and end positions of the slit are noted with the letters 'S' and 'E', respectively, in each $x-t$ map.

Dual flow or spicules with relatively low LOS velocity are observed as broadening of the H$\alpha$ line profile in both blue wing and red wing (refer Figure \ref{Fig3} RP no. 10--12), whereas for the upflows or the downflows, the broadening is seen either in the red wing or in the blue wing of the line profiles.

\subsection{Upflows} 
\begin{figure}
\centering
	\includegraphics[width=0.210\textwidth]{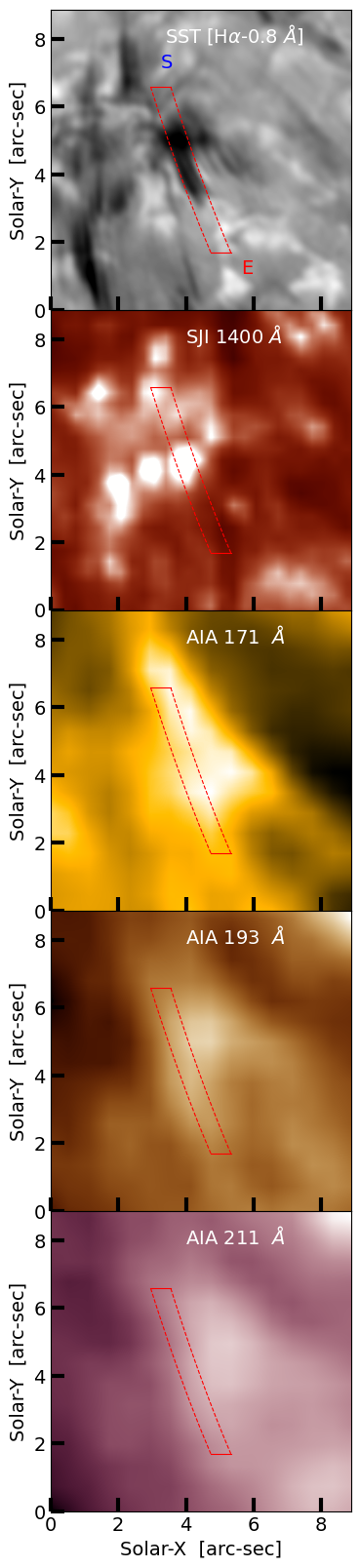}
     \includegraphics[width=0.257\textwidth]{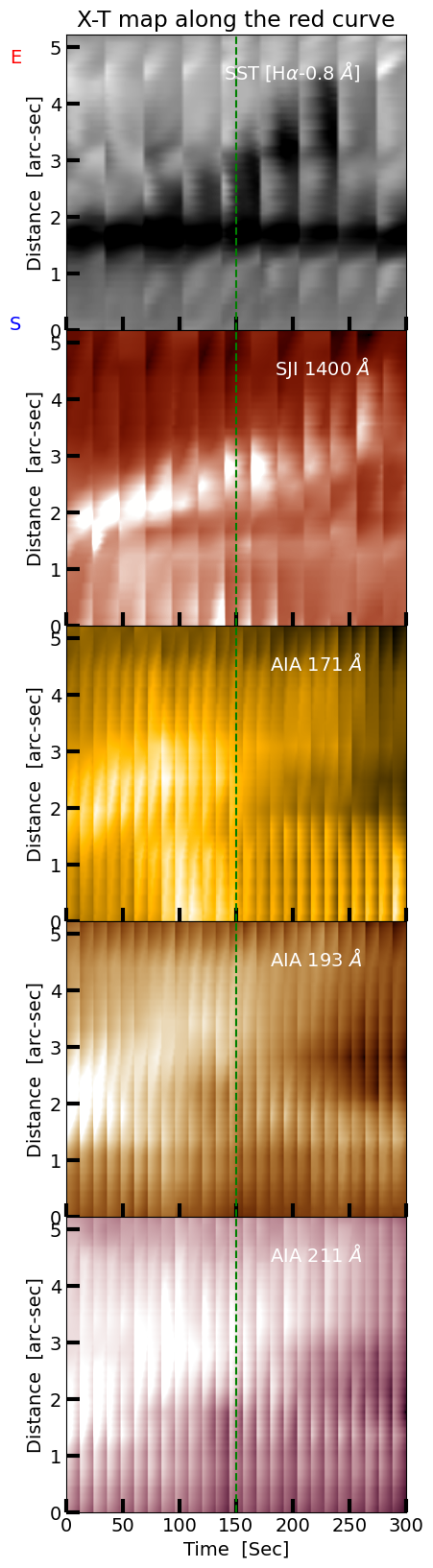}
	\caption{The space-time map at various passbands sampling different heights in the solar atmosphere. The red contour over the background images in the left column represents the path over which the space-time plot is drawn; S and E indicates starting and end positions of the slit. The right column represents the $x-t$ map of SST/H$\alpha$-0.8\angs, SJI 1400\angs, AIA 171\angs, AIA 193\angs and AIA 211\angs from top to bottom, respectively. The green dashed line indicates the time corresponding to the image displayed in the left panel. An animation of the SST/H$\alpha$-0.8\angs sequence illustrating this event is available. The animation provides a clear depiction of the evolution of a spicule through the virtual slit. These spicules are observed to move away from the magnetic flux concentration region, indicative of upflows in the chromosphere. The animation comprises a total of 8 frames spanning time from 08:26:29 to 08:30:29 UT. The real-time duration of the animation is 0.4 seconds.}
\label{Fig5}
\end{figure}

\begin{figure}
\centering
  \includegraphics[width=0.194\textwidth]{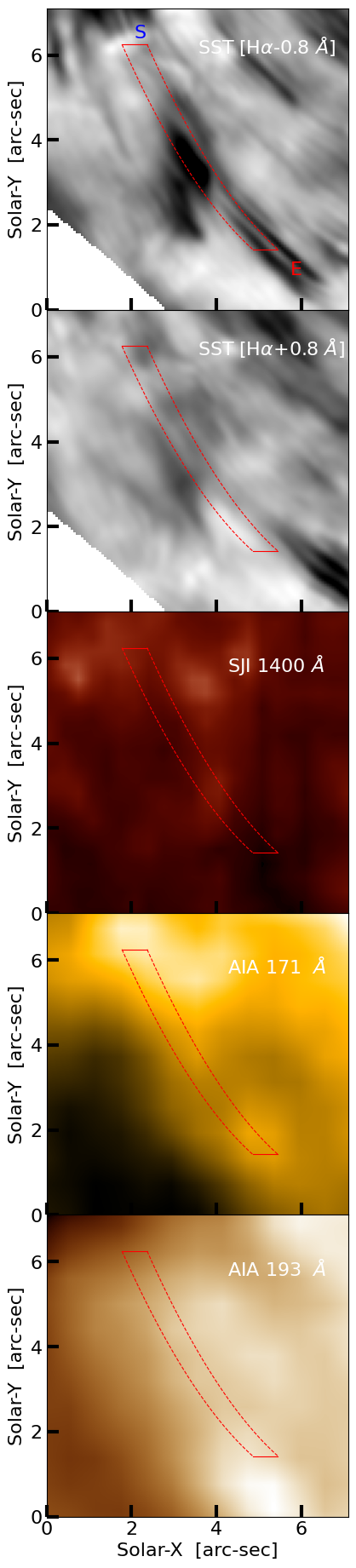}
  \includegraphics[width=0.277\textwidth]{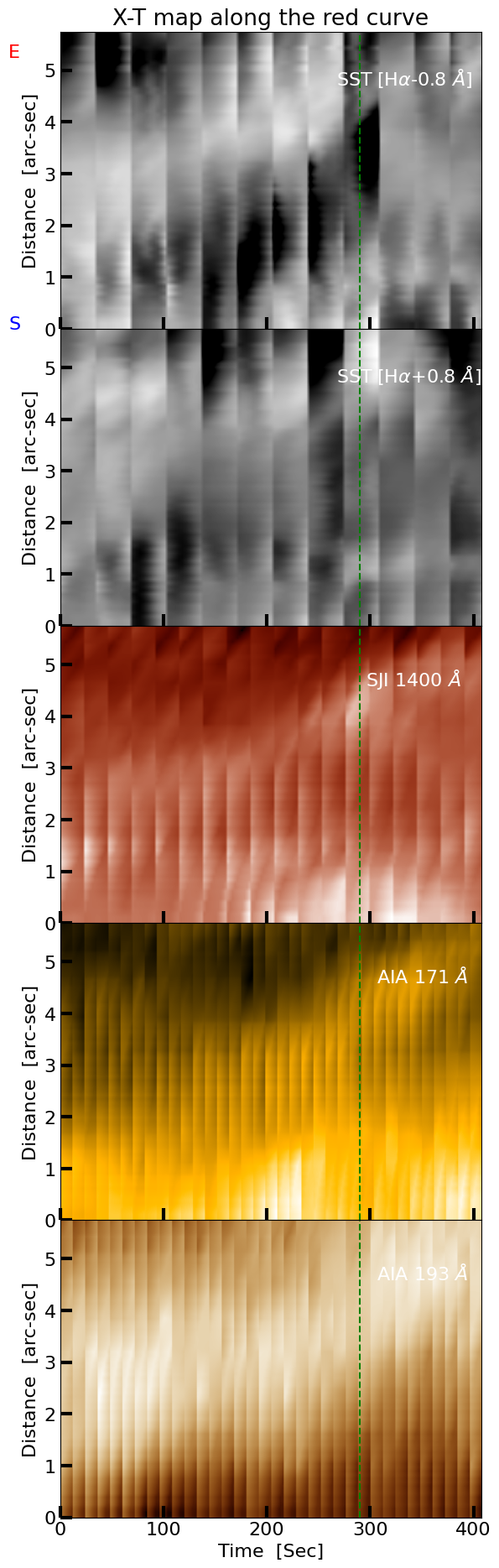}
  \caption{Same as Figure \ref{Fig5}, however this also includes SST/H$\alpha$ red wing images. The red wing images (H$\alpha$+0.8\angs) also shows upflow, which indicates the RPs corresponding to this show broadening in the red wing too, but representing a net upflow in the solar atmosphere. The animation depicting this event in the blue wing of H$\alpha$ shows a clear illustration of the evolution of a spicule through the virtual slit, although the seeing condition was not stable during this event. The animation consists of a total of 12 frames spanning time from 08:04:48 to 08:11:05 UT. The real-time duration of the animation is 0.6 seconds.}

\label{Fig6}
\end{figure}

Figures \ref{Fig5} and \ref{Fig6} show $x-t$ maps of upflowing REs along the path marked by the red contour over the background images shown on the left side of each Figure. The left side images show almost a one-to-one correspondence of the features between the blue/red wing images from SST/H$\alpha$ and the IRIS/SJI 1400\angs. The absorption features in the H$\alpha$ images appear as brighter regions in the SJI 1400\angs. Conversely, magnetic flux concentrations appear bright in all the channels from the chromosphere to the corona.

The corresponding $x-t$ maps show that the absorption features in the blue/red wing of the H$\alpha$ spectral line move upward and become visible as emission features at TR and coronal passbands viz. SJI 1400\angs, AIA 171\angs, AIA 193\angs and AIA 211\angs respectively. These $x-t$ maps indicates that the upflowing features (REs) are promptly heated to the TR and coronal temperatures. This suggests that the upflowing plasma is multi-thermal in nature. This becomes more evident when observing a time series of images (a animation).

Unlike Figure \ref{Fig5}, which shows upflowing pattern in blue wing images, Figure \ref{Fig6} shows similar upflowing pattern in both blue and red wing images. This indicates, the REs shown in Figure~\ref{Fig6} exhibit a broadened spectral line profiles with a net upflow/blueshift.  

\subsection{Downflows} 
\begin{figure}
\centering
\includegraphics[width=0.20\textwidth]{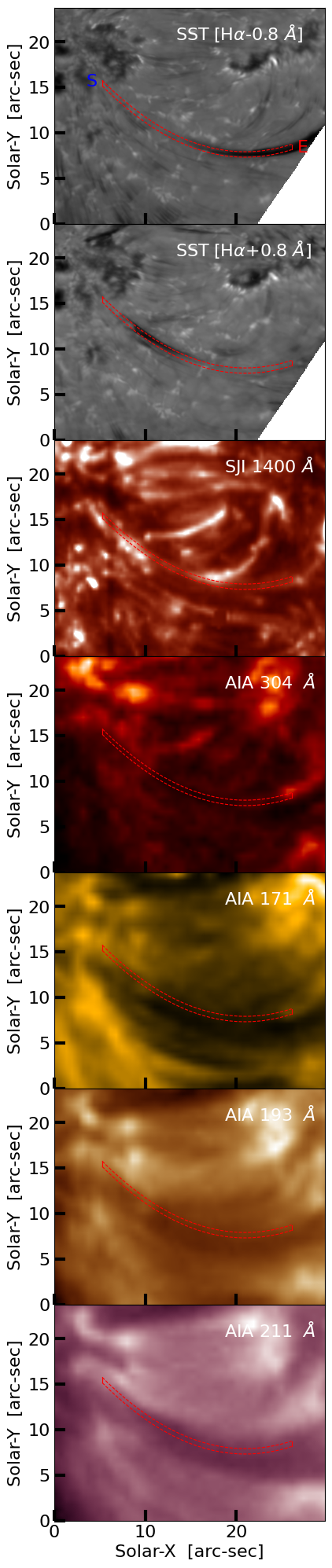}
\includegraphics[width=0.199\textwidth]{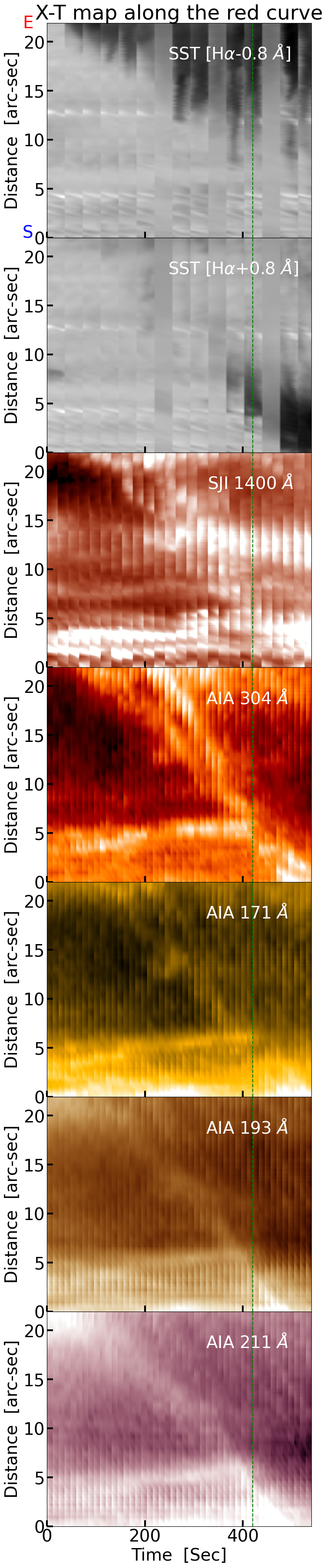}
\caption{Same as Figure~\ref{Fig5} for a downflowing rapid excursion first seen in the blue wing and later (around 400 sec) seen in red wing of the H$\alpha$. This indicates conversion of RBE to RRE. The enhanced emission seen in TR and coronal passbands almost parallel to the time axis could be due to the continuous emission of stationary plasma. However, the morphology of this feature is different from the on-disk spicules; this could be a low-lying loop structure. The animation depicting the H$\alpha$ event provides a clear illustration of the evolution of a curved feature originating from outside the FoV and subsequently entering the dark pore region, as observed in the blue, followed by the red wing of H$\alpha$. The animation consists of a total of 15 frames spanning time from 08:42:12 to 08:50:48 UT. The real-time duration of the animation is 0.75 seconds.}

\label{Fig7}
\end{figure}

Figure~\ref{Fig7} shows similar $x-t$ maps for the downflowing feature. Here it is observed that the downflowing RBE converting to downflowing RRE at a later time (about 300~s). These REs show stronger excursion (absorption) as compared to other REs with a higher LOS velocity.  
There is also an enhanced emission parallel to the time axis seen in the AIA passbands, predominantly in the AIA 304\angs, but the corresponding feature is not seen in H$\alpha$ images. This could be due to the continuous emission of the stationary plasma corresponding to coronal temperature. 

\begin{figure}
\centering
    \includegraphics[width=0.184\textwidth]{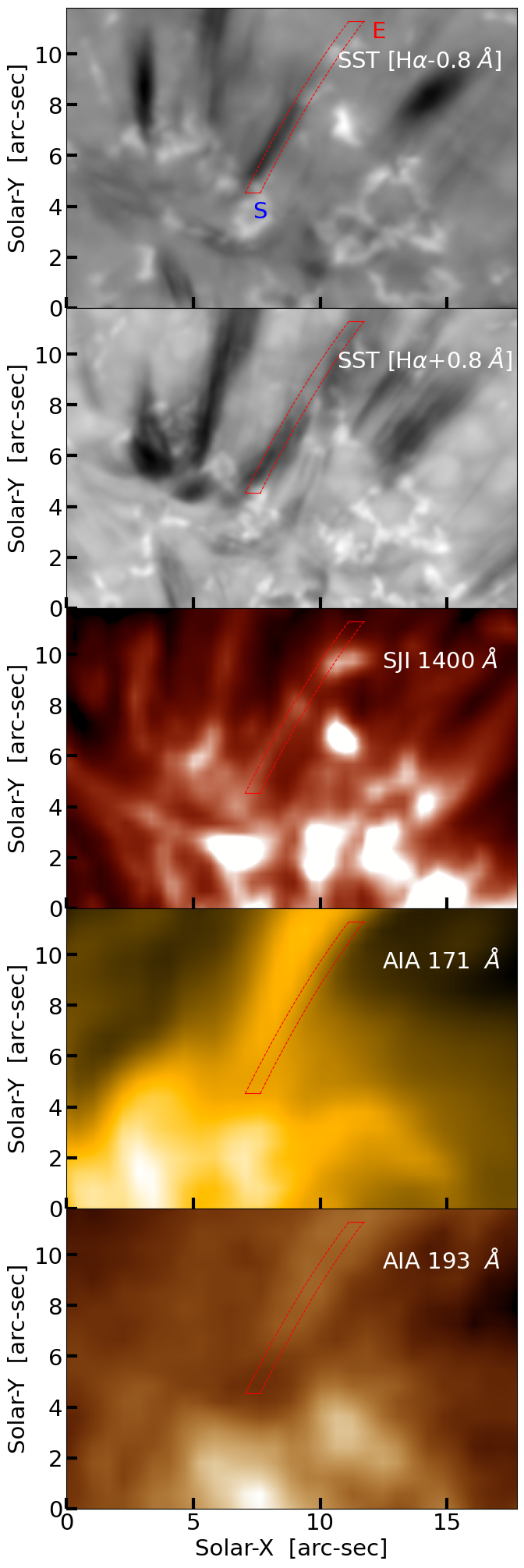}
    \includegraphics[width=0.27\textwidth]{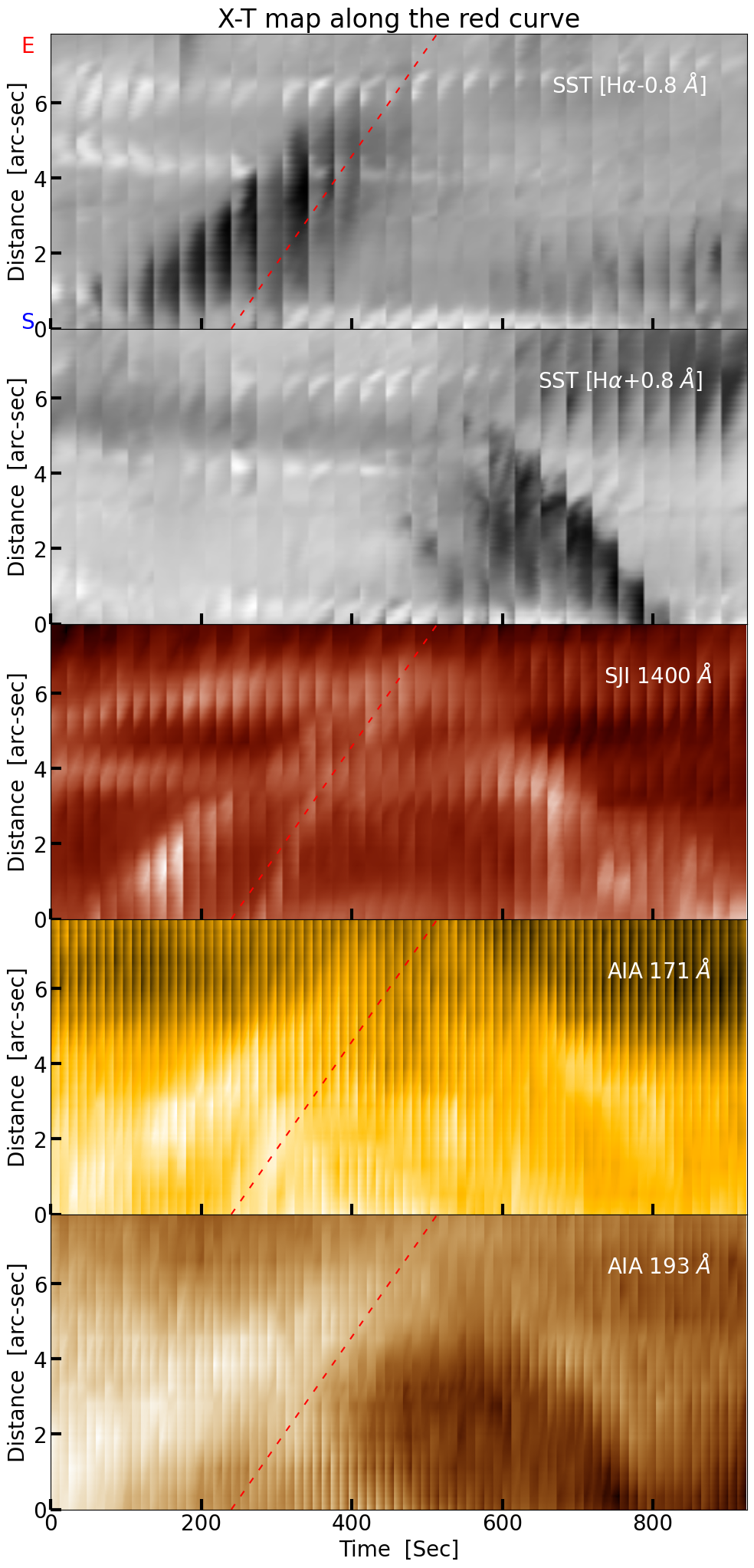}
\caption{Space-time plots of the observed RBEs/RREs at various temperature passbands sampling different heights in the solar atmosphere. The top two images show the upward and downward motion one after the other, and the remaining subimages show the parabolic path observed in TR and coronal passbands. The upward velocity is relatively lower than the downward velocity. It is to be noted that the upflow comprises of two distinct upward flows that are clearly discernible in the TR and coronal passbands. The red dashed line represents the secondary upflow. The animation depicting this event in  H$\alpha$, AIA 171\angs, and AIA 193\angs distinctly shows the evolution of a spicule through the virtual slit, along with several other spicules observed in both the blue and red wings. These spicules exhibit movement away from the magnetic flux concentration, followed by motion towards it, essentially demonstrating clear upflow and downflow phenomena in the chromosphere. The comparison of AIA sequence with H$\alpha$ sequence reveals the heating spicular top to coronal temperature. The H$\alpha$ portion consists of 24 frames while the AIA portion uses 71 frames to span from approximately 08:07 to 08:22 UT. The total real-time duration of this animation is 3.55 seconds.}

\label{Fig8}
\end{figure}

\subsection{Sequential up and downflows} 
\begin{figure}
\centering
    \includegraphics[width=0.197\textwidth]{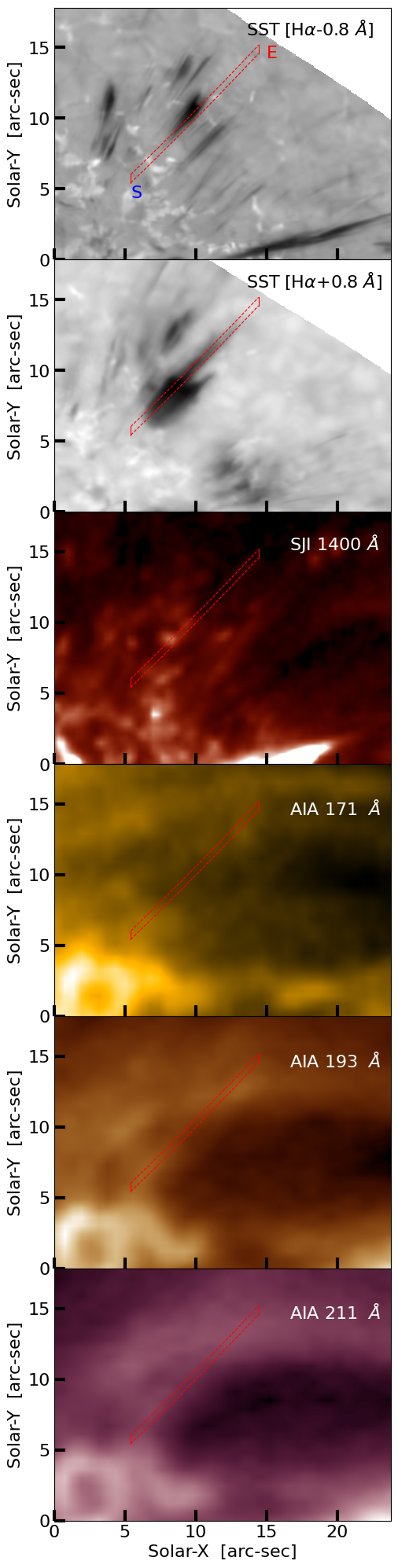}
    \includegraphics[width=0.275\textwidth]{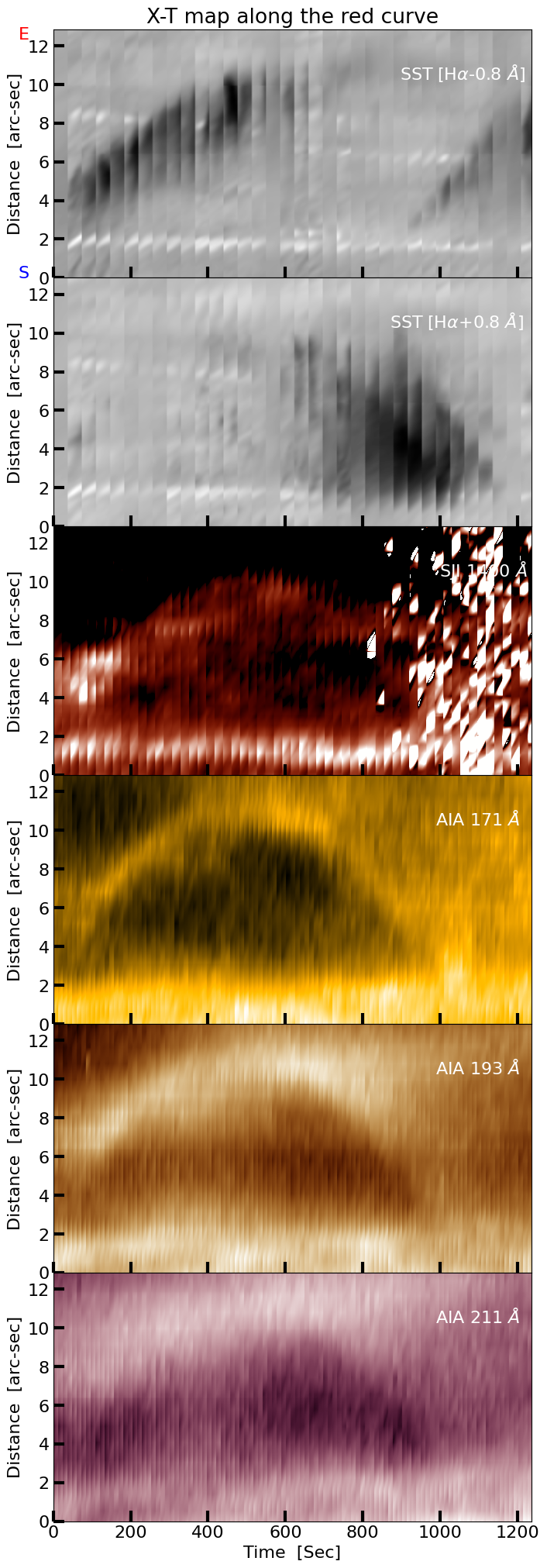}
\caption{Same as Figure \ref{Fig8} for some other location of RBE/RRE, which shows parabolic path in TR and coronal passbands. Figure \ref{DEM_map} shows the DEM analysis of this event.
An animation depicting this event in H$\alpha$, AIA 171\angs, AIA 193\angs and 211\angs is available. The H$\alpha$ sequence consists of 26 frames while the AIA portions have 93 frames. The animation spans the time from about 08:49 to 09:08 UT. The H$\alpha$ sequence was sped up by a factor of 4 to match the AIA. The real-time duration of the animation is 5.15 seconds.}
\label{Fig9}
\end{figure}

We have also observed upflows and downflows along the same path one after the other in different channels of the solar atmosphere. Figures~\ref{Fig8} and \ref{Fig9} depict such flows. From the blue wing and red wing image of SST/H$\alpha$ (left panel of Figure~\ref{Fig8}) it is clear that the upflows originate from the magnetic flux concentration region and descend towards it. These upflows and downflows take on the appearance of a loop-like structure when observed in the AIA passbands. This becomes more evident when viewing the accompanying animation. The space-time plots provide a clear illustration of a simultaneous upflow observed across all passbands, covering the chromosphere to the corona, starting at the same time and reaching a maximum height of approximately 6~Mm in nearly 10~mins. 

The upflow (in Figure~\ref{Fig8})  comprises of two distinct upward flows that are clearly discernible in the TR and coronal passbands. Similarly, there are two secondary downflows located in proximity to the primary downward flow, but they are relatively weak and only single downflowing feature is seen in the TR and coronal passband images. As the upflow reaches its maximum height, the downflow phase becomes apparent in all passbands, resulting in a parabolic path in the spacetime plot. The upward phase persisted for a longer duration (approximately 10~mins) compared to the downflowing phase (around 4 mins). Consequently, the upward velocity (about 10.5\kms) was relatively lower than the downward velocity (14.0\kms) in the sky plane. The entire journey of the plasma during the upflows and downflows spans a duration of 14.5 mins.

Figure \ref{Fig9} exhibits a similar flow pattern, featuring a distinct parabolic path in the space-time plot. Here, H$\alpha$ red wing images show two downflowing features; however, their combined signature appears as a single feature in the AIA and SJI passbands. Similar to previous observations, the upward velocity measures 13 km s$^{-1}$, which is less than the downward velocity of 19 km s$^{-1}$, in the sky plane.

\subsection{Simultaneous up and down flows}

In addition to the sequential upward and downward flows along the same path, a more intricate plasma motion occurs where both upward and downward flow exist at the same time and at the same spatial location. This differs from the sequential flow pattern, where downflow commences only after the complete termination of upflow. Observing such flows can be challenging due to their complex nature. Since there are flows in both directions at the same spatial location, it results in a broadening in both wings of the H$\alpha$ spectral line, having a relatively stronger excursion either in the red or in the blue. These profiles belong to RP numbers 10, 11, and 12 as shown in Figure \ref{Fig3}. 

\begin{figure}
\centering
    \includegraphics[width=0.25\textwidth]{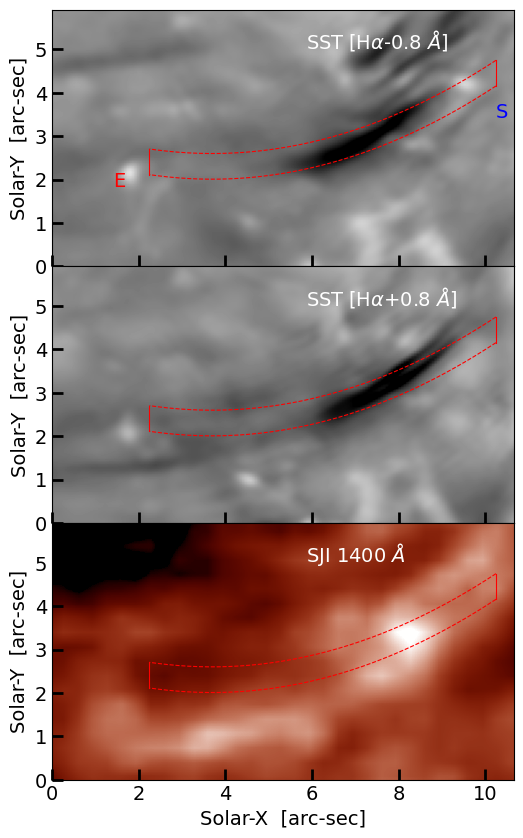}
    \includegraphics[width=0.22\textwidth]{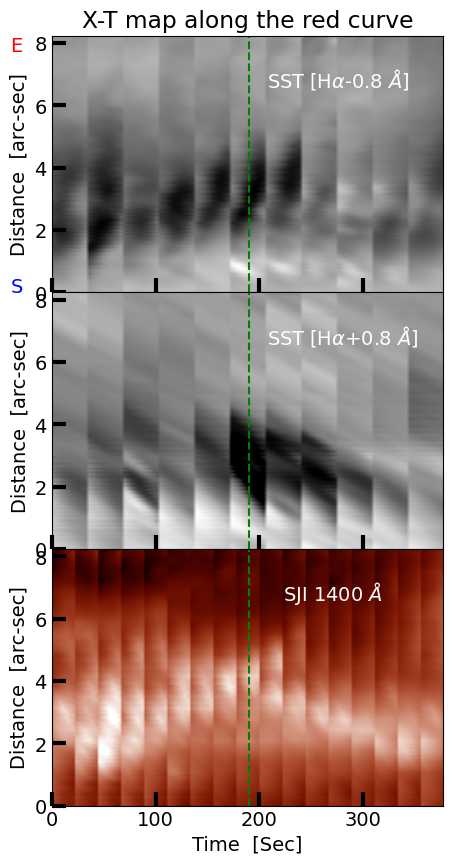}
\caption{The space-time plot of dual flows observed at chromosphere and TR temperature regimes. The X--type signature seen in the SJI 1400\AA~$ x-t$ plot around 180~s indicates the both upflow and downflow at the same time along the same path. The green dashed line indicates the time corresponding to the image displayed in the left panel. The mosaic representation of this event is shown in Appendix \ref{Fig_10_Mosaic} to provide a more clear picture of the simultaneous flow. An animation depicting the H$\alpha$ event is available. The animation shows two panel corresponding to blue (upper) and red (lower) wing of H$\alpha$. The comparison of the two panel shows a presence of upflow and downflow at the same around 180 s. The animation consists of 11 frames spanning time from 08:04:48 to 08:10:32 UT. The real-time duration of the animation is 0.55 seconds.}

\label{Fig10}
\end{figure}

\begin{figure}
\centering
    \includegraphics[width=0.201\textwidth]{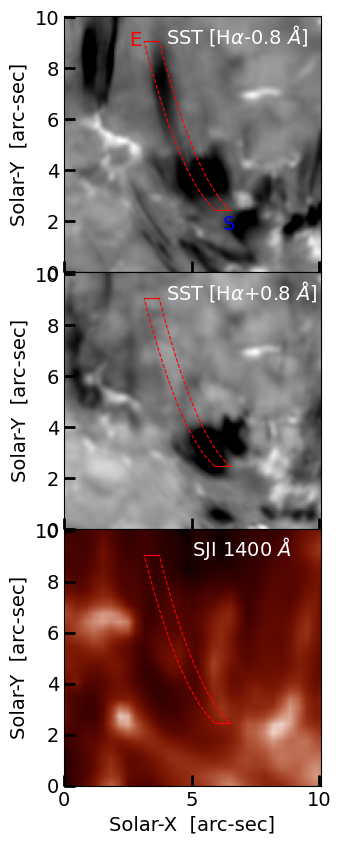}
    \includegraphics[width=0.25\textwidth]{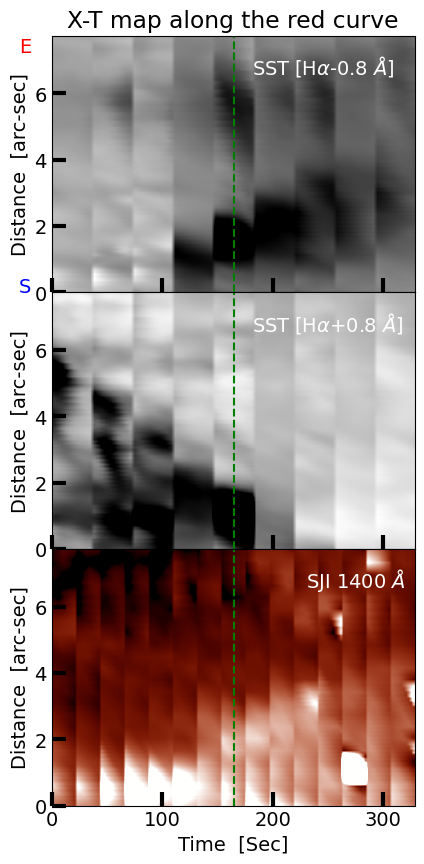}
\caption{Same as Figure~\ref{Fig10}, however, the x-type signature seen in the SJI 1400\AA~ images is not clear here. Refer to \ref{Fig_11_Mosaic} for the mosaic representation of this event. An animation depicting the event consists of 9 frames spanning time from 08:58:45 to 09:04:30 UT. The real-time duration of the animation is 0.45 seconds.}

\label{Fig11}
\end{figure}

Figure \ref{Fig10} depicts the signatures of the dual/simultaneous flows (co-existence of upflow and downflow spatio-temporally) along a spicular path in the chromosphere and TR. The left panel of the Figure shows the presence of an absorption feature in both wings of the H$\alpha$ spectral line at the same spatial location. Similar to other examples, this absorption feature appears as an emission feature with some offset, but a significant portion is within the virtual slit. From the $x-t$ map, it appear that the downflow has already started before the upflow terminates completly. The upflow is observed during 20--230~s, whereas the downflow is observed around 170~s onwards. Both upflow and downflow can be seen around 4 arcsec and at 180 sec. This is clearly visible in the TR channel as an {\bf "X"} mark at around 180~s. However, such features are not observed in coronal channels. Similar characteristics of the plasma flow have been shown in Figure~\ref{Fig11}. However, the signature of a downflow in TR is not as prominent as observed in Figure~\ref{Fig10} but, it can be seen around 6 arcsec with reduced contrast. We show mosaics in appendix \ref{appA} to present a more clear picture of such kinds of flows in the chromosphere. The mosaic is shown only for those frames (for Figure \ref{Fig10}; frame No: 4--6 and for Figure \ref{Fig11}; frame No:3--5) where there is a simultaneous flow.

\subsection{Signature of downflows in Transition Region}

The spectral lines corresponding to TR temperature (4 $\times$ 10$^{4}$ -- 2.2 $\times$ 10$^{5}$ K) show a predominance of redshift/downflow. To study these plasma flows at TR temperature, we used IRIS Si IV 1394\angs ($\sim$ 10$^{4.8}$ K) spectra.

\begin{figure*}
\centering
	  \includegraphics[width=0.90\textwidth]{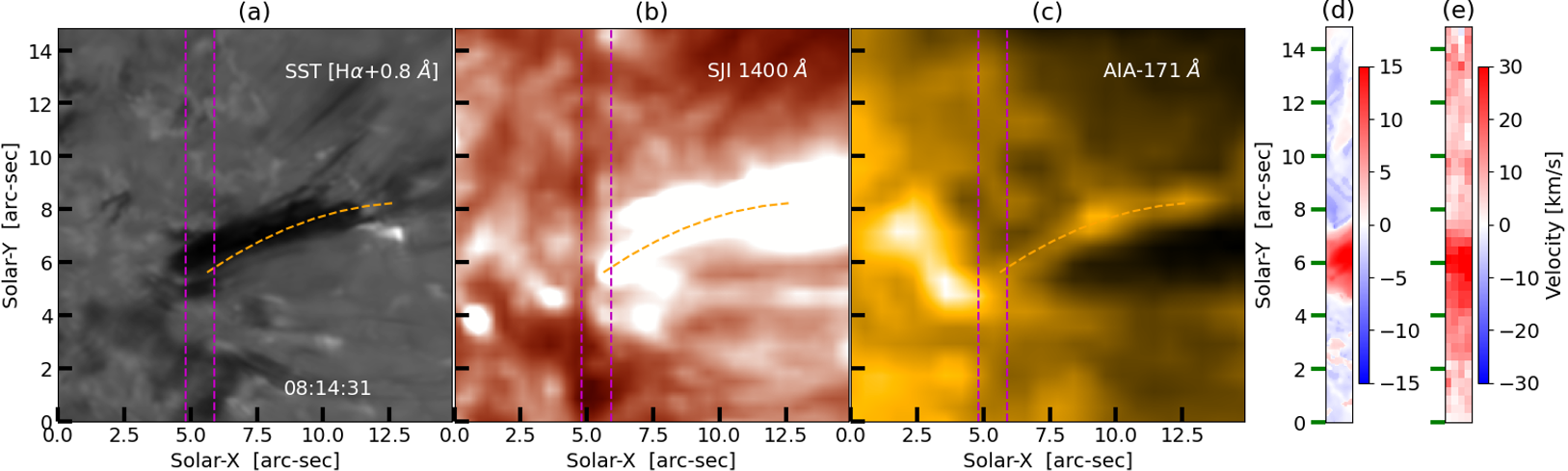}
   
\caption{
The figure displays a snapshot of the H$\alpha$+0.8\angs, TR (Si 1400\angs), and corona (AIA 171\angs), with each of these regions exhibiting a common feature highlighted by an orange dashed line. The purple parallel lines represent the coverage of the slit, spanning approximately 1 arcsec. The accompanying sub-images on the extreme right show Doppler maps corresponding to the region covered by the slit both in Chromosphere and in TR. These Doppler maps show dominant redshifts at the locations corresponding to the downflowing feature seen in H$\alpha$-0.8\angs. A animation depicting this event in the red wing of H$\alpha$ clearly shows the bundles of spicules going into the pore region. The animation consists of 25 frames spanning time from 08:07:41 to 08:21:23 UT. The real-time duration of the animation is 1.25 seconds.}
\label{Fig12}
\end{figure*}

\begin{figure*}
\centering
	  \includegraphics[width=0.90\textwidth]{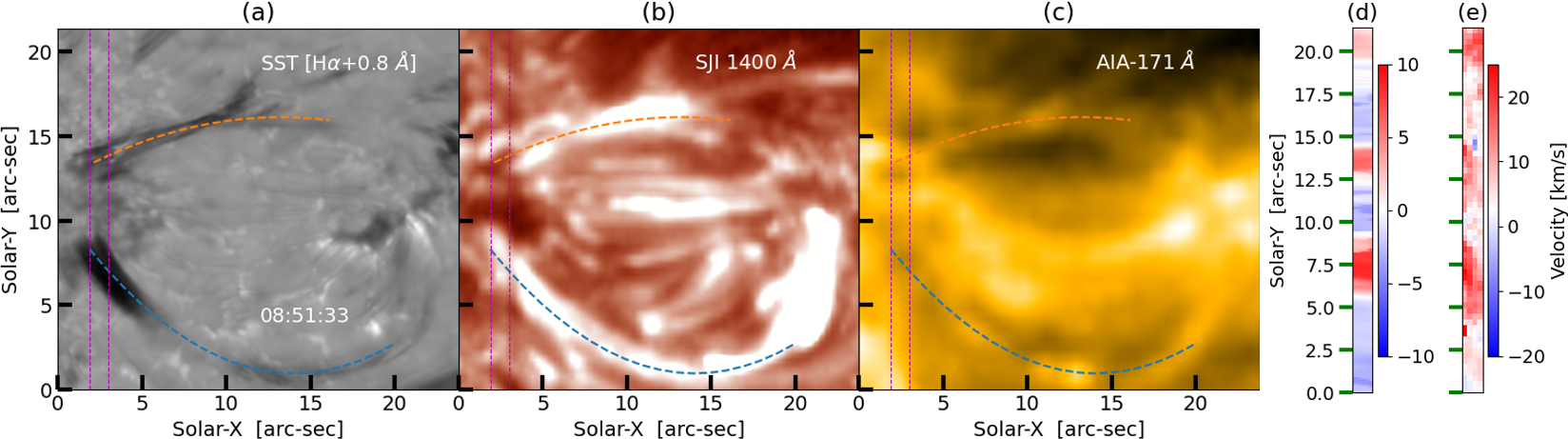}
    
\caption{Same as Figure \ref{Fig12} at a later time. Here two such features coincide with the slit. The corresponding Doppler maps obtained from Chromosphere and TR observations show predominant downflows. A animation depicting this event in the red wing of H$\alpha$ clearly shows two features going into the dark pore region. The animation consists of 35 frames spanning time from 08:48:31 to 09:09:04 UT. The real-time duration of the animation is 1.75 seconds.}
\label{Fig13}
\end{figure*}

During the observation, all (4) IRIS slits were aligned with the tip of the termination of the downflowing feature, as shown in Figure~\ref{Fig12}a. The accompanying animation gives a clear impression of a common feature (dashed orange line) observed throughout the solar atmosphere sampling from thousands to million-degree corona, which is moving downward in the solar atmosphere in the form of bundle of spicules. The right panels (d and e) of the Figure~\ref{Fig12} display the chromosphere and TR Doppler velocities obtained from SST/H$\alpha$ and IRIS data (Si IV 1394\angs) respectively, for all the slit positions covering an approximately one arcsec FoV. We found similar Doppler velocities demonstrating a similar behaviour in other spectral lines such as C II 1336\angs, Si IV 1403\angs, and Mg II k 2796\angs apart from Si IV 1394\angs sampling a temperature regime correponding to transition region. However, only one Doppler velocity map (Si IV 1394\angs $\sim$ 10$^{4.8}$ K ) is shown here. The comparison of the snapshot, the accompanying animation, and the Doppler map reveals that the dominant redshifts/downflows observed in the Doppler map are located at the same spatial positions where the downflowing feature terminates. Figure~\ref{Fig13} further illustrates this phenomenon, showing two downflowing features (indicated by the orange and blue dashed lines), both of which produce a dominant redshift in the Doppler velocity maps.

\section{Summary and Discussion} \label{sec5}

The discovery of type II spicules \citep{de2007tale} has generated much interest in the scientific community as they are considered potential candidates to explain observed emissions in TR and corona. 
Several studies have provided their evidence that type II spicules, or REs, are rapidly heated to temperatures corresponding to the transition region and corona \citep{2009ApJ...701L...1D, 2009ApJ...707..524M, 2014ApJ...792L..15P, 2015ApJ...799L...3R, 2011Sci...331...55D, 2016ApJ...820..124H, 2016ApJ...830..133K, 2019Sci...366..890S}. Although the role of spicules is well-established in the dynamics of the transition region, their impact on the corona and solar wind remains a subject of debate (as discussed in \cite{2019ARA&A..57..189C}). 

We studied the impact of REs and similar spectral features, which correspond to different temperature regimes in the solar atmosphere, using coordinated observation from ground-based and space-based instruments mentioned in section~\ref{sec2}. We used the terms REs and spicules interchangeably to collectively represent RBE, RREs, and dual excursion RPs, as the two are the same features. The data clearly shows the dominance of REs in the chromosphere when observed at the wings of the H$\alpha$ spectral line. The accompanying animation clearly shows that most of the features seem to move away from the magnetic flux concentration region when viewed in blue wing images \citep{2009ApJ...705..272R}. Conversely, they tend to move towards magnetic flux regions when observed in red wing images (as described in \cite{2021A&A...647A.147B}).
It is worth noting that some of the RBEs exhibit downflowing behaviour, although not as prominently as the upflowing RBEs. Similarly, RREs can display upflowing behaviour, though they are not as dominant as downflowing RREs. This behaviour can be attributed to the complex motion exhibited by spicules \citep{2013ApJ...769...44S, 2014Sci...346D.315D}.

Our study begins with detecting RBE/RREs and similar spectral features using $k-$means clustering from the H$\alpha$ data since we were interested in studying the impact of these features in the solar atmosphere. We used $k-$means clustering to find the spatial location that belongs to RBEs, RREs and the dual excursion RPs in the whole FoV. It is observed that number of RPs that show RBEs is always less than the RPs that represent RREs, which indicates that RREs shows a wider range of velocities than RBEs. Other than these RPs that correspond to RBEs and RREs alone, there exist a set of RPs that show broadening on either wing (RP numbers 10, 11, and 12 of the Figure~\ref{Fig3}). This kind of REs were overlooked earlier, assuming that these RPs did not represent net flows and were merely the result of thermal broadening (\cite{2022MNRAS.509.5523N}). However, our region of $x-t$ plots (refer Figure~\ref{Fig10} and \ref{Fig11}), corresponding to these RPs, show spatio-temporal occurrence of the upflow and downflow in the solar chromosphere. These RPs tend to occur in conjunction with traditional RBE/RREs, which is evident in Figure \ref{Fig4}, as one can see that the grey part (RPs 10-12) appears close to the blue and red part demonstrating different distribution of LOS velocity. In addition to this, these RPs can also be associated with a simultaneous upward and downward motion of one spicule/bundle of spicule in the solar chromosphere. It is important to note that while these flows are detectable up to the transition region, they remain concealed in coronal passbands may be due to the limited resolution. Therefore, the significance of dual excursion RPs should not ignored.

Figures~\ref{Fig5} and \ref{Fig6} depicting upflowing REs provide clear evidence that these spicules rapidly heat to temperatures in the transition region and corona, is in agreement with \cite{2019ARA&A..57..189C} [and references therein]. Moreover, these upflowing signatures coexist at various temperatures, ranging from the chromosphere to the corona, displaying a multi-thermal nature. As time progresses, these spicules appear to move upward in the solar atmosphere, showcasing their dynamic behaviour. These REs are rooted near or from the magnetic flux concentration when observed in the H$\alpha$ wing images. REs appear to show a multi-thermal behaviour, especially if we assume ionisation equilibrium. It is possible that non-equilibrium ionisation affects REs due to the fast dynamics and steep temperature gradients. In such cases, ionisation and recombination could be delayed given a temperature change in REs, which could lead to features being visible in more spectral lines than those given by the ionisation equilibrium temperature. Non-equilibrium effects could narrow down the temperature range where REs exist, but will not significantly alter our analysis. Regardless of the ionisation state, our observations are in line with a scenario where heating frequently occurs, typically concentrated towards the top of the spicule. In these examples of upflowing REs, Figure~\ref{Fig5} represents entirely upflowing RBEs, adhering to the traditional behaviour of RBEs without showing any significant contribution in the red wing. However, Figure~\ref{Fig6} reveals similar upflow pattern in its red wing in addition to the blue wing $x-t$ plot, indicating a net upflow despite the broadening of the spectral line profile. The presence of nearly cospatial RBE and RRE could be due to the presence of multiple spicule threads that have different amounts of twists along with the upward movement of whole spicule. Therefore, dual excursion RPs may be associated with simultaneous upward and downward flows, only upflows, or downflows, depending on the complex motion exhibited by the spicules. This result aligns with findings by \cite{2022MNRAS.515.2672V}, who identified both net upflows and downflows associated with dual excursion RPs.

In contrast to the conventional RREs, some RREs exhibit plasma motion in the opposite direction, moving toward the magnetic flux concentration regions as detailed in \cite{2021A&A...647A.147B}. We have identified and selected these features using CRISPEX \citep{2012ApJ...750...22V} as the k-means clustering only gives the spatial location of REs and not the direction of plasma flow associated with the REs. The space-time map (Figure~\ref{Fig7}) clearly illustrates the flow of plasma from higher atmospheric layers to lower ones. Similar to the upflows, the downflowing material exhibits multi-thermal characteristics. It is important to note that the downflowing plasma initially appears as an RBE and then, as time progresses, transforms into an RRE. A closer look at the sub-images of this feature in different passbands (left Figure \ref{Fig7}) indicates that the morphology of this feature is different from the typical on-disk spicules. This could possibly be a low-lying loop as it looks too curved and much longer than a typical on-disk spicule. The conversion from RBE to RRE could be due to the LOS projection velocities along the loop. It was detected by k-means clustering because these features exhibited comparable spectral signatures across the disk \citep{2018A&A...611L...6P}.

Furthermore, we observed sequential upflow and downflow along the same trajectory, illustrating a parabolic path in the space-time map as noted by \cite{2014ApJ...792L..15P} and \cite{2015ApJ...806..170S}. However, these previous studies focused solely on the transition region (TR). In contrast, our results illustrate similar behaviour across multiple coronal passbands and offers the first distinct instance of the spicule parabola in AIA 211 Å. Although $x-t$ maps corresponding to Figures \ref{Fig8} and \ref{Fig9} clearly show the upflow followed by the downflow, it remains challenging to ascertain whether the material returning to lower heights during the drainage is the same or different from the material that ascended previously, as the draining is happening after the complete termination of upflow. The accompanying H$\alpha$ wing image animations of these two events reveal a unique characteristic of these spicules: not only does the spicule within the virtual slit descend, but essentially all spicules in the field descend almost simultaneously and such collective behaviour is seldom seen. This behaviour differs significantly from off-limb spicules, where various spicules ascend and others descend simultaneously, displaying a mixed behaviour. The AIA sequences in Figure \ref{Fig9} distinctly portray a moving blob in all three AIA filters above the spicule's top which is a coincidence, but consistent with the idea of hot front of the spicule. These AIA animations strongly imply that the features observed in AIA are associated with the spicule. Conversely, to verify that these emissions indeed stem from coronal temperatures rather than the TR temperature, we present $x-t$ maps of the Differential Emission Measure (DEM) of the same event in various temperature ranges (see Appendix \ref{appB}). The DEM images also depict a clear parabolic trajectory akin to Figure \ref{Fig9}, strongly indicating that the features observed in AIA are indeed heated to coronal temperatures.

In addition to this, a notable observation is that spicules tend to vanish when they reach the maximum height of their journey in the H$\alpha$ wing images but reappear during the descending phase after travelling some distance towards the lower atmosphere. However, in the transition and coronal passbands, REs persist throughout their journey. The disappearance from the H-alpha wing image could be attributed either to the conversion of most of the plasma to higher temperatures (beyond the temperature of the line formation of H$\alpha$) or due to low projected velocity. Additionally, the later part of the spicule parabola (downward-flowing spicules) exhibits an increase in opacity before seemingly disappearing from the red-wing H$\alpha$ images. This suggests the cooling of spicules in the chromosphere as they are drained back and could result in the chromospheric heating. The average lifetime (approx. 15 min) in our case was found to be greater than the average lifetime mentioned by \cite{2015ApJ...806..170S} for the whole event. These downflows are believed to be associated with the drainage of excess plasma carried by the spicules back to the chromosphere (\cite{1982ApJ...255..743A}; \cite{2021A&A...654A..51B}). This drainage is necessary because the mass flux associated with these features is several orders of magnitude higher than that of the solar wind (\cite{1977A&A....55..305P}). \cite{2012ApJ...749...60M} found signatures of downflowing patterns in the relatively cool emission channels of the lower solar corona that were speculated to be returning heated spicule plasma. Interestingly, in our analysis, the sky-plane velocity associated with the upflow was lower than the downward velocity, which suggests the influence of solar gravity. Similar to the upflow and downflow, these flows also exhibit a multi-thermal nature, indicating the coexistence of plasma at various temperatures.

In addition to these findings, the prevalence of redshifts (with occasional hints of blueshifts) observed in spectral lines of the TR remains a topic of intense debate, with several theories attempting to elucidate this observed shift (\cite{1993ApJ...402..741H}; \cite{1977A&A....55..305P}; \cite{1982ApJ...255..743A}; \cite{hansteen2010redshifts}). Despite the predominance of downflows observed in the IRIS spectral line corresponding to TR temperature \cite{1997SoPh..175..349B}, \cite{1999ApJ...522.1148P}, and more recently, \cite{2018ApJ...859..158S} demonstrated that most high-speed quiet sun TR downflows typically vanish at chromospheric temperatures. It has been proposed that the absence of downflow signatures in the chromosphere might stem from the breakdown of these downflows upon reaching chromospheric temperatures, given that the plasma density in the chromosphere is several orders of magnitude higher than in the TR and the corona. 

Our results (Figures \ref{Fig12} and \ref{Fig13}) offer insights into the observed shift in the IRIS TR temperature spectral lines. We present two downward-moving features in H$\alpha$ wing images that exhibit a predominant redshift and correlate directly with the redshifts observed in the TR. Figure \ref{Fig12} illustrates the descent of bundles of spicules (marked by an orange dashed line) from coronal passbands down to the chromosphere, including the TR. Meanwhile, Figure \ref{Fig13} displays two downward-moving features: the first (marked by an orange dashed line) corresponds to the one depicted in Figure \ref{Fig12} but at a later time, while the second feature (marked by a blue dashed line) resembles the structure shown in Figure \ref{Fig7}, resembling a low-lying loop yet displaying a spectral signature akin to RBE/RREs over the disk \citep{2018A&A...611L...6P}. This confirms that at least some of the downward moving material from the coronal heights is in the form of spicules or spicular-like spectra are responsible for producing observed redshifts in TR temperature spectra as suggested by \cite{1982ApJ...255..743A}, \cite{1977A&A....55..305P} and \cite{2021A&A...654A..51B}. Moreover, these types of downward-moving features, characterised by strong redshifts in the TR temperature spectral line, indeed have chromospheric counterparts. The comparison of the Doppler map of Figure \ref{Fig12} and \ref{Fig13} shows that the strong redshifts in the TR do have a chromospheric counterpart. On the contrary, alongside the strong spatio-temporal redshifts/downflows in the TR and corona, we also observe some weak redshifts in a TR temperature spectral line (Si IV 1394 Å) with no corresponding signatures in the chromosphere. 
and suggests that the plasma flow corresponding to TR temperature may not extend to the chromospheric height, resulting in distinct flows observed in the chromosphere independent of the TR temperature plasma flow in the Doppler velocity map. Our findings align with the hypothesis of the breakdown of weak downflows as they reach chromospheric temperatures.

\section{Conclusions} \label{sec6}

We focused on examining the response of the TR and corona to the RE/spicules observed in the chromosphere. Our findings clearly show that spicules are rapidly heated to TR and coronal temperatures and persist throughout their trajectory, indicating a multi-thermal nature. Although there is a clear spicule parabola exceeding 10 MK temperature, our study cannot definitively confirm the heating of the spicules to this temperature range.  Additionally, the cooling of spicules in the chromosphere as they are drained back could result in the chromospheric heating. However, a definitive conclusion can only be reached if it is established that the loss of the spicular energy is significantly consumed into the chromospheric heating. Further investigation in this area could provide valuable insights into spicule heating temperatures and the chromospheric heating. Furthermore, we hypothesize that the weak contrasting flow observed in the TR and chromosphere may result from the breakdown of the TR flow upon reaching chromospheric temperatures. Nevertheless, further studies in this area can offer a clearer understanding of the cospatial redshift/downflows in TR and the chromosphere.

\begin{acknowledgments}
We thank the referee for invaluable suggestions and comments. The Swedish 1--m Solar Telescope is operated on the island of La Palma by the Institute for Solar Physics of Stockholm University in the
Spanish Observatorio del Roque de los Muchachos of the Instituto de Astrofísica de Canarias. IRIS is a NASA small explorer mission developed and operated by LMSAL with mission operations executed at NASA Ames Research Center and major contributions to downlink communications funded by ESA and the Norwegian Space Centre. SDO is a mission for NASA Living With a Star program. The
SDO/HMI data were provided by the Joint Science Operation Centre (JSOC). This work has been supported by the Research Council of Norway through its Centre of Excellence scheme, project number 262622.
\end{acknowledgments}

\section*{Data Availability}
 
The observational data utilized in this study from AIA and HMI onboard SDO are available at http://jsoc.stanford.edu/ajax/lookdata.
html and data from IRIS mission are available at https://iris.lmsal.com/search/. 
SST observations were obtained from https://dubshen.astro.su.se/sst$\_$archive/search.

\newpage
\appendix

\section{Mosaic of Figure 10 and 11}  \label{appA}

\begin{figure}
\centering
\includegraphics[width=0.48\textwidth]{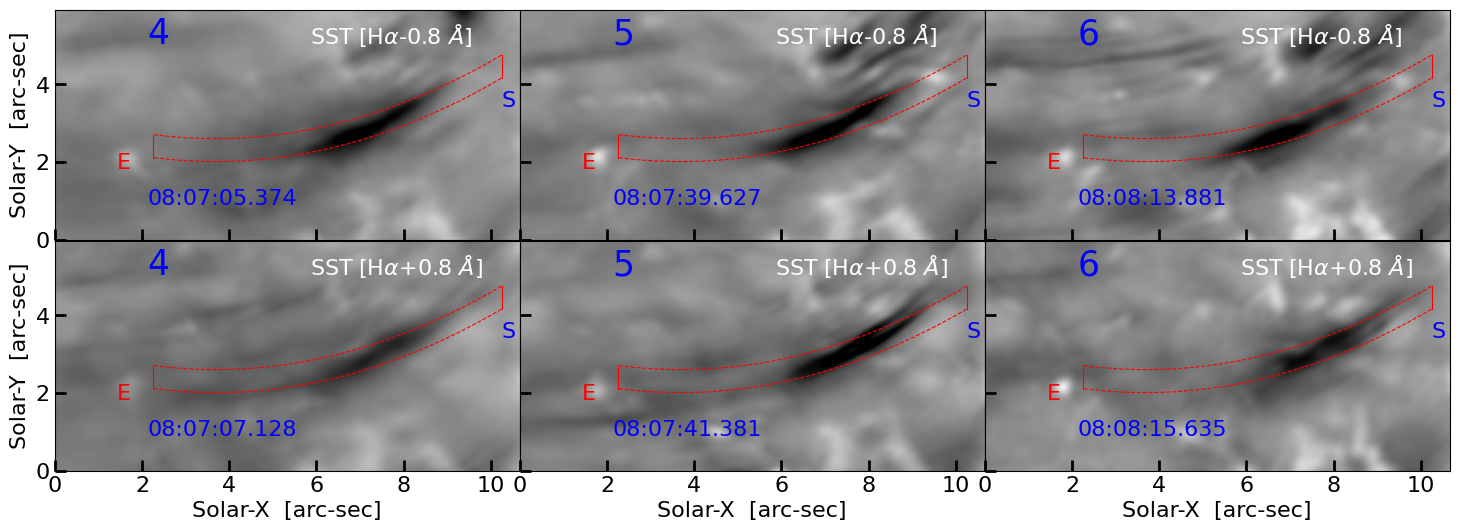}
\caption{A mosaic comprising frames 4 to 6 of Figure \ref{Fig10} illustrates simultaneous upflow and downflow.}
\label{Fig_10_Mosaic}
\end{figure}

\begin{figure}
\centering
\includegraphics[width=0.48\textwidth]{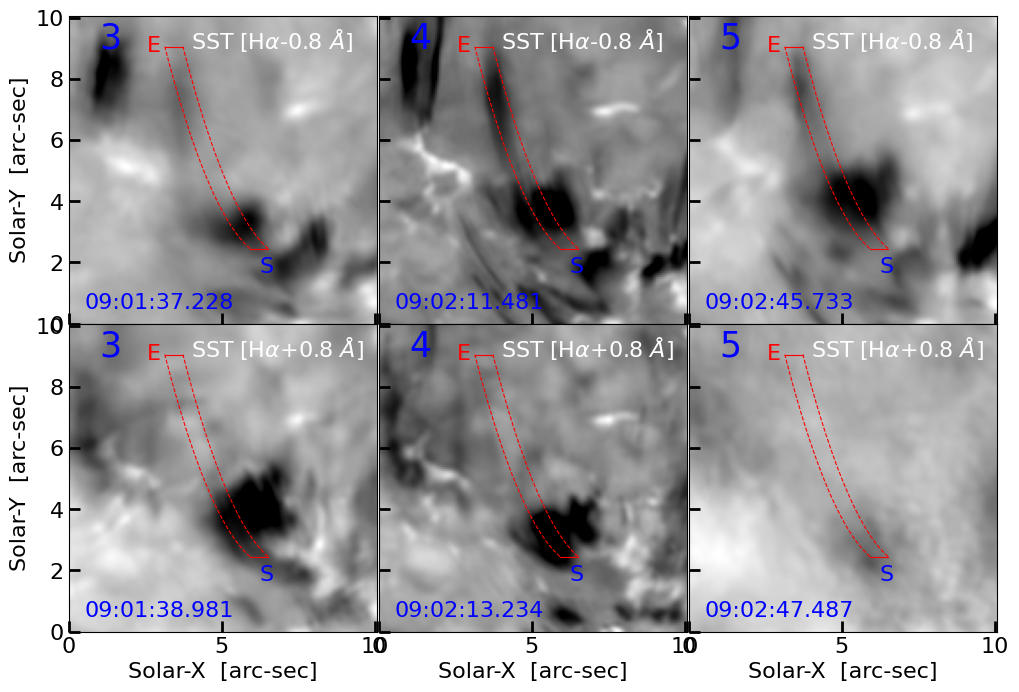}
\caption{Similar to \ref{Fig_10_Mosaic}, a mosaic of Figure \ref{Fig11} depicts frames 3 to 5, showcasing simultaneous upflow and downflow. Due to seeing, the spicule bundle is washed out (as seen in the lower right panel).}
\label{Fig_11_Mosaic}
\end{figure}

Figures \ref{Fig_10_Mosaic} and \ref{Fig_11_Mosaic} depict mosaic representations of simultaneous upflow and downflow in the solar chromosphere. In Figure \ref{Fig_10_Mosaic}, the comparison of blue and red wing images of H$\alpha$ reveals a single spicule moving upward and downward almost simultaneously. This might arise from multiple spicular threads with similar morphology positioned one above the other along the LOS and moving in opposite directions. Conversely, Figure \ref{Fig_11_Mosaic} demonstrates the simultaneous motion of a group of spicules in the solar chromosphere.

\section{Differential Emission Measure (DEM) Analysis}  \label{appB}

\begin{figure}
\centering
\includegraphics[width=0.42\textwidth]{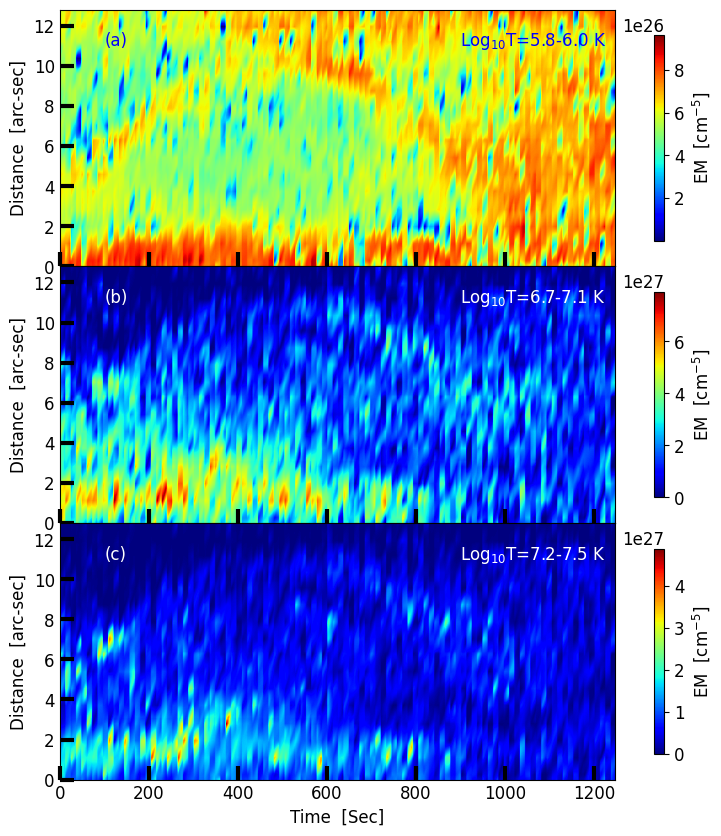}
\caption{The space-time map of DEM of the event shown in Figure \ref{Fig9} at different temperature regimes: (a) \(\log_{10} T = 5.8 \text{ -- } 6.0 \, \text{K}\), (b) \(\log_{10} T = 6.7 \text{ -- } 7.1 \, \text{K}\), and (c) \(\log_{10} T = 7.2 \text{ -- } 7.5 \, \text{K}\) clearly showing a parabolic path.}
\label{DEM_map}
\end{figure}

We employ a modified version of the sparse inversion DEM\footnote{\url{http://paperdata.china-vo.org/yang.su/DEM/sparse_em_v1.001_ys.zip}} code (Older; \citep{2015ApJ...807..143C}; Modified -- \citep{2018ApJ...856L..17S}) to interpret the temperature contributions in the AIA FoV for each EUV channel to the spicule (Figure \ref{Fig9}). This modified version incorporates improved settings, which enhance the constraints on plasma Emission Measure (EM) at high temperatures, leading to more accurate DEM results. The DEM ($\xi$(T)) quantifies the amount of plasma emitting within a specific temperature range and is linked to the electron density of the plasma (n$_e$) and the column height along the LOS (dz), expressed as \(\xi(T) = \int_{0}^{\infty} n_e^2(T) \, dz\). Figure \ref{DEM_map} shows three \(x\)-\(t\) maps from the DEM analysis of the event shown in Figure \ref{Fig9} at different temperature regimes, corresponding to (a) \(\log_{10} T = 5.8 \text{ -- } 6.0 \, \text{K}\), (b) \(\log_{10} T = 6.7 \text{ -- } 7.1 \, \text{K}\), and (c) \(\log_{10} T = 7.2 \text{ -- } 7.5 \, \text{K}\). These $x-t$ maps distinctly illustrate a spicule parabola, providing strong evidence of the spicule's heating to coronal temperatures. We cannot definitively confirm the heating of the spicule to temperatures exceeding 10 MK as obtained from DEM analysis. However, our findings provide compelling evidence of the spicule's heating to coronal temperatures throughout its entire journey.

\newpage
\bibliography{sample631}{}
\bibliographystyle{aasjournal}

\end{document}